\RequirePackage{fix-cm}
\documentclass[smallextended]{svjour3_Preprint}       
\smartqed  
\usepackage{graphicx}
\usepackage{mathptmx}      
 \journalname{Experiments in Fluids}
\usepackage{siunitx}
\usepackage{upgreek}
\usepackage{color}
\usepackage[T1]{fontenc}
\usepackage{natbib}

\newcommand{\Prt}{\mathit{Pr}_\mathrm{t}}
\newcommand{\Ma}{\mathit{Ma}}
\newcommand{\Mai}{\mathit{Ma}_\mathrm{i}}
\newcommand{\MaiLES}{\mathit{Ma}_{\mathrm{i},\mathrm{LES}}}

\newcommand{\Rei}{\mathit{Re}_\mathrm{i}}
\newcommand{\ReiLES}{\mathit{Re}_{\mathrm{i},\mathrm{LES}}}
\newcommand{\Retheta}{\mathit{Re}_\theta}
\newcommand{\Res}{\mathit{Re}_S}
\newcommand{\ResLES}{\mathit{Re}_{S,\mathrm{LES}}}
\newcommand{\St}{\mathit{St}}

\newcommand{\Sct}{\mathit{Sc}_\mathrm{t}}

\newcommand{\xs}{\mathit{x}/\mathit{S}}
\newcommand{\ys}{\mathit{y}/\mathit{S}}

\newcommand{\xsimp}{\mathit{x}_\mathrm{imp}/\mathit{S}}
\newcommand{\ximp}{\mathit{x}_\mathrm{imp}}

\newcommand{\fracAirVol}{{\varphi}_\mathrm{air}}
\newcommand{\fracHe}{\mathit{Y}_\mathrm{He}}
\newcommand{\fracHeVol}{{\varphi}_\mathrm{He}}
\newcommand{\particleDensity}{\mathit{\rho}_\mathrm{p}}
\newcommand{\particleCount}{\mathit{N}_\mathrm{p}}
\newcommand{\fracHeU}{\overline{\fracHe'u'}/u_\infty}
\newcommand{\fracHeV}{\overline{\fracHe'v'}/u_\infty}
\newcommand{\molarRatio}{m}
\newcommand{\molarRatioDefinition}{{M_\mathrm{Air}}/{M_\mathrm{He}}}

\newcommand{\TR}{\mathit{T}_{0,\mathrm{i}}/\mathit{T}_{0,\infty}}

\newcommand{\relaxTime}{\tau_\mathrm{p}}
\newcommand{\relaxLength}{l_\mathrm{p}}

\renewcommand{\u}{\overline{u}/u_\infty}

\newcommand{\uv}{\overline{u'v'}/u_\infty^2}

\newcommand{\mm}{\mathrm{mm}}
\newcommand{\um}{\mathrm{\upmu m}}
\newcommand{\us}{\mathrm{\upmu s}}
\newcommand{\ns}{\mathrm{ns}}
\renewcommand{\deg}{^\circ}

\newcommand{\caseExpRef}{{E-I}}
\newcommand{\caseExpShock}{{E-II}}
\newcommand{\caseNumRef}{{N-I}}
\newcommand{\caseNumShock}{{N-II}}

\begin{document}

\title{Experimental investigation of the turbulent Schmidt number in supersonic film cooling with shock interaction
}

\titlerunning{Exp. invest. of the turb. Schmidt number in supersonic film cooling with shock interaction}        

\author{Pascal Marquardt         \and
        Michael Klaas \and
				Wolfgang Schr\"oder
}


\institute{Pascal Marquardt \at
              Chair of Fluid Mechanics and Institute of Aerodynamics Aachen\\
							W\"ullnerstr. 5a \\
							52062 Aachen \\
							Germany\\
              Tel.: +49-241-8090420\\
              Fax: +49-241-8092257\\
              \email{p.marquardt@aia.rwth-aachen.de}           
           \and
           Michael Klaas \at
              Chair of Fluid Mechanics and Institute of Aerodynamics Aachen\\
							W\"ullnerstr. 5a \\
							52062 Aachen \\
							Germany
					 \and
           Wolfgang Schr\"oder \at
              Chair of Fluid Mechanics and Institute of Aerodynamics Aachen\\
							W\"ullnerstr. 5a \\
							52062 Aachen \\
							Germany	
}

\date{05.11.2019}

\maketitle

\begin{abstract}
The interaction of an impinging shock and a supersonic helium cooling film is investigated experimentally by high-speed particle-image velocimetry. A laminar helium jet is tangentially injected into a turbulent air freestream at a freestream Mach number $\Ma_\infty=2.45$. The helium cooling film is injected at a Mach number $\Mai=1.30$ at a total temperature ratio $\TR=0.75$. A deflection $\beta=8\deg$ generates a shock that impinges upon the cooling film. A shock interaction case and a reference case without shock interaction are considered. The helium mass fraction fluctuations are measured and the turbulent mass flux as well as the turbulent Schmidt number are determined qualitatively.
 For comparison, large-eddy simulation (LES) results of a comparable flow configuration are used.
The streamwise and wall-normal turbulent mass fluxes are in qualitative agreement with the LES data. 
The turbulent Schmidt number differs significantly from unity. Without shock interaction, the turbulent Schmidt number is in the range $0.5 \leq\Sct\leq 1.5$ which is in agreement with the literature.
With shock interaction, the turbulent Schmidt number varies drastically in the vicinity of the shock interaction. Thus, the experimental results confirm the numerical data showing a massively varying turbulent Schmidt number in supersonic film cooling flows, i.e., the standard assumption of a constant turbulent Schmidt number is valid neither without nor with shock interaction.
\end{abstract}

\section{Introduction}
\label{sec:Introduction}

\begin{figure}[b]
	\centering
		\includegraphics{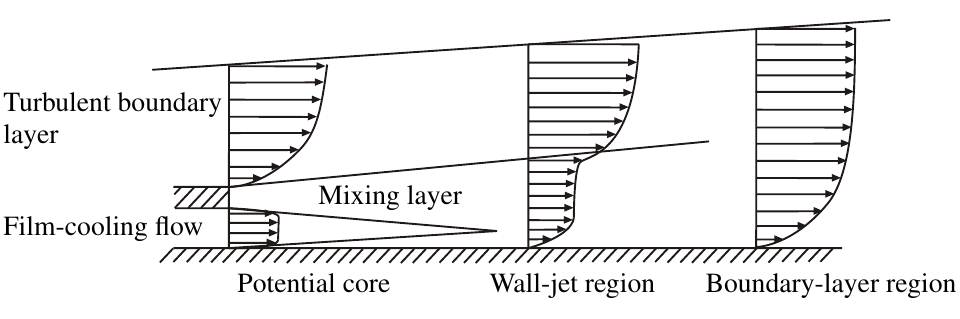}
	\caption{Flow schematic with velocity profiles indicating three distinct flow regions (\cite{Seban62,Juhany94c}) in a tangential film-cooling configuration (\cite{Konopka2012}).}
	\label{fig:flow_schematic_nachKonopka}
\end{figure}

In supersonic applications with high thermal loads, e.g., scramjet combustors, where the gas temperature exceeds the thermal limit of the surface material, so-called film cooling, i.e., the injection of cold gas along the surface, is often used as a cooling method. 
As shown in the review of \citet{Goldstein71}, a promising cooling concept for supersonic flows is a cooling configuration with tangential injection. 
A schematic drawing of this cooling concept beneath a turbulent boundary layer is shown in figure~\ref{fig:flow_schematic_nachKonopka}. According to \citet{Seban62} and \citet{Juhany94c}, the flow can be divided into three main regions. 
The first region is the potential core right downstream of the injection, which is bounded by the mixing layer that emanates from the lip and the slot flow boundary layer. The potential core ends where the mixing layer and the slot-flow boundary layer merge. At this location, the wall-jet region, which is characterized by intense mixing, starts. Further downstream, the flow relaxes to an undisturbed turbulent boundary layer which is denoted as boundary-layer region. Shock waves that might be present in the flow field can cause shock induced separation of the cooling film. 
If the onboard fuel, i.e., hydrogen, is considered as cooling fluid, the flow field exhibits a significant density gradient in the mixing layer. The combination of a free shear layer, a mixing layer, a wall bounded jet, a boundary layer, strong density and species gradients, and shock induced separation results in a complex flow field.
To design efficient scramjet engines that consume as little coolant as possible, a precise prediction of the complex flow and temperature field and the distribution of coolant species is essential. This requires a profound understanding of the turbulent mixing mechanisms within the cooling-film flow, especially when flow separation occurs.

Cost efficient numerical methods to predict the flow in scramjet combustors, i.e., Reynolds-averaged Navier-Stokes (RANS) simulations, face the problem of modeling the turbulent transport in the flow. Common models, e.g., the k-$\epsilon$ model (\cite{Jones72}) or the k-$\omega$ model (\cite{Wilcox88}), assume a constant relationship between the turbulent diffusion of momentum and the turbulent diffusion of heat and mass. The ratio between the momentum eddy diffusivity and the heat eddy diffusivity is denoted as turbulent Prandtl number $\Prt$ and the respective ratio for the mass eddy diffusivity is called the turbulent Schmidt number $\Sct$.
When simulations consider the temperature as a passive scalar, the turbulent Prandtl and the turbulent Schmidt number are identical.
The present study is focused on the turbulent mass transport. 
Hence, the following review is limited to studies concerning the turbulent transport of mass or of a passive scalar.
In general, a constant turbulent Schmidt number, which is at least problem dependent, is assumed in RANS simulations.
In reviewing the optimal turbulent Schmidt numbers for engineering flow fields relevant to atmospheric dispersion, e.g., jet-in-cross flows and plume dispersion in boundary layers, \cite{Tominaga2007} found a range for the optimal turbulent Schmidt number, i.e., the turbulent Schmidt number that showed the best agreement with experimental results, in the range $0.2\leq \Sct \leq 1.3$. Therefore, depending on the flow problem, a wide range of turbulent Schmidt numbers must be considered in RANS simulations. The strong effect of the turbulent Schmidt number on RANS of scramjet related flows was shown by \cite{Eklund2001}, who performed RANS simulations of an ethylene-fueled scramjet combustor with turbulent Schmidt numbers in the range $0.2\leq\Sct\leq1.0$. The turbulent Schmidt number strongly affects the location of the pre-combustion shock train, the peak pressure, the ignition of the primary fuel, and the amount of heat release within the combustor.

The turbulent Schmidt number not only depends on the flow problem, but also varies locally within the flow field. \cite{He99} performed RANS simulations of a jet-in-cross flow at various turbulent Schmidt numbers and compared the results to experimental results of \cite{Crabb1981} and \cite{Kamotani1974}. Even though the authors found the best agreement between the numerical and the experimental results at a turbulent Schmidt number of $\Sct=0.2$, they concluded that a variable Schmidt number may be needed especially for low momentum flux ratio jet-in-cross flows. Later, \cite{Jiang2009} analyzed the temperature distribution within a general combustor experimentally and numerically. They found an optimal agreement at a turbulent Schmidt number of $\Sct=0.5$. Nevertheless, the authors pointed out, that the concept of calculating turbulent scalar transfers based on the modeled momentum transfer and a turbulent Schmidt number is limited and should be improved and new approaches should be developed.
\cite{Brinckman2007} proposed a scalar-variance model which predicts a variable turbulent Schmidt number. They found that the scalar-variance model showed better agreement with experimental data over a range of reacting and non-reacting flows compared to computations for fixed $\Sct$. Another eddy-viscosity-based turbulence model with variable $\Sct$ was proposed by \cite{Goldberg2010}. For various benchmark flows including a scramjet combustor flow the new turbulence model resulted in a better agreement with experimental data compared to simulations with fixed $\Sct$.

In the past decade, numerical results which resolve all relevant flow scales, i.e., direct numerical simulations (DNS), became available for reasonably high Reynolds numbers. \cite{Li2009}, \cite{Wu2010}, \cite{Araya2012}, and \cite{Li2016} performed DNS of spatially developing incompressible boundary layers including a passive scalar up to Reynolds numbers $\Retheta=2300$. The simulations show variations of the turbulent Schmidt number within the boundary layer. The peak values close to the wall were in the range $1.1\leq\Sct\leq1.9$. Further off the wall, the turbulent Schmidt number drops below unity. 
%
%
%
%

The experimental determination of the turbulent Schmidt number and the turbulent scalar fluxes is an extremely challenging task, especially in high-speed flows.
Different measurement techniques have been used in low-speed water flow. \cite{Koochesfahani2000} performed molecular tagging velocimetry (MTV) and laser induced fluorescence (LIF) measurements of a mixing layer between two water streams. They used a pulsed laser grid to excite a phosphorescent tracer compound and recorded the displacement of the excited molecules within a given time interval. Additionally, the laser excites fluorescence of fluorescein tracers in one of the streams to capture the tracer concentration. They demonstrated the ability to determine the velocity-concentration correlation, i.e., the turbulent scalar fluxes.
\cite{Hjertager2003} combined particle-image velocimetry (PIV) and planar laser induced fluorescence (PLIF) measurements in a confined wake flow. They captured the planar distribution of the mean concentration, concentration fluctuations, and the turbulent scalar fluxes. For measurements of the scalar transport in water flows, simultaneous PIV/PLIF measurements have been established as the standard measurement technique.

In air flows, which are of interest when the phenomenon of shock/boundary-layer interaction plays a role in the film cooling setup, the fluorescent dyes that are normally used in water, e.g., Rhodamine 6G or fluorescein, cannot readily be used. Additionally, the significantly lower density reduces the fluorescence emission intensity. Thus, instantaneous concentration measurements in air flows are considerably more challenging than measurements in water, especially in high-speed flows where the density may be lower by yet another order of magnitude.
\cite{Melnick2014} performed simultaneous PIV measurements and flow visualization of a subsonic boundary layer at Reynolds numbers in the range $2100 \leq \Retheta \leq 8600$. They applied light seeding in the freestream and added dense seeding in the boundary layer through a slit in the wall. By using the intensity of a subsampled version of the particle image, the authors were able to calculate the correlation between the velocity and the smoke intensity. They found correlation coefficients between the velocity deficit and the smoke intensity as high as $0.7$. However, they did not estimate any turbulent scalar fluxes.
Recently, \cite{Combs2019} successfully applied simultaneous PIV and PLIF on a supersonic boundary layer at a Mach number $\Ma=5$. They used naphtalene as tracer for the PLIF measurement which was added to the boundary layer on the bottom wall. The authors determined the mole fraction of the tracer with a measurement uncertainty of $\pm20\%$.
Additionally, they determined the turbulent scalar fluxes.

Concerning supersonic film cooling, the literature on turbulent transport of species is very limited.
Early studies on shock/cooling-film interaction mainly focused on the determination of the wall heat flux. Investigations of \cite{Alzner71}, \cite{Kamath90}, \cite{Holden90} and \cite{Olsen90} led to different conclusions about the extent of the shock induced increase of the peak heat transfer. \cite{Juhany94c} concluded that the differences in the literature are related to the flow region where the shock impinges upon the cooling film.
Measurements with regard to the use of hydrogen fuel as coolant were conducted by \cite{Alzner71}. The authors performed measurements with air and hydrogen injection on an axisymmetric model at a freestream Mach  number $\Ma_\infty=6$. They found that at hydrogen injection, considerably less injected mass is required for a similar cooling effect. However, the majority of the experimental investigations use helium injection to generate the density gradient between the freestream and the cooling film. \cite{Kwok91} used a sampling probe to measure the helium concentration of a tangential $\Mai=1.78$ helium injection into a $\Ma_\infty=3.0$ freestream. By analyzing profiles of the mean helium fraction, they found that most of the mixing occurs in the top third of the mixing layer. However, the sampling probe was not able to measure the instantaneous helium concentration. 
\cite{Konopka2013a} performed LES of film cooling with laminar injection of helium and hydrogen beneath a turbulent boundary layer. The injection and the freestream Mach number were $\Mai=1.30$ and $\Ma_\infty=2.44$.
The authors showed that a shock that impinges upon the cooling film drastically decreases the cooling effectiveness compared to a no-shock cooling flow. 
The shock induced separation bubble at hydrogen injection is about $23\%$ larger compared to the helium injection. Furthermore, the separation bubble at hydrogen injection was found to be very stable. Thus, the cooling effectiveness is reduced by $40\%$ at helium injection and by $30\%$ at hydrogen injection.
Downstream of the separation bubble, the turbulent species flux off the wall is almost doubled.
Additionally, the authors investigated the distribution of the turbulent Schmidt number. They found the turbulent Schmidt number to vary across the shear layer in the range $0.5 \leq \Sct \leq 1.5$.

In summary, for supersonic film cooling, the turbulent mass flux is essential when a light weight gas, i.e., hydrogen, is considered as coolant to reduce the required coolant mass flow rate.
Since the mixing of the cooling film and the freestream strongly influences the cooling effectiveness, an efficient film cooling design requires accurate predictions of the turbulent mass flux, and hence, the turbulent Schmidt number. Especially when shock waves impinge on the cooling film, the turbulent mixing is increased and the importance of accurate prediction of the turbulent transport is even more pronounced. Nevertheless, in standard turbulence models a constant turbulent Schmidt number is assumed. 
While \cite{Konopka2013a} investigated the turbulent mixing of hydrogen and helium cooling films including the effect of shock interaction numerically by LES, there is no experimental data of the turbulent flux and the turbulent Schmidt number of a supersonic film cooling flow without or with shock impingement available.
Hence, it is the scope of this study to measure the turbulent mass transport of a supersonic film cooling configuration to experimentally substantiate the findings of \cite{Konopka2013a}. This means, the distribution of the turbulent mass fluxes as well as the distribution of the turbulent Schmidt number are investigated for film cooling flows without and with shock interaction.

 The experimental setup that was previously used for investigations with isothermal (\cite{Marquardt2019a}) and cooled (\cite{Marquardt2019b}) air injection is extended to use helium as cooling fluid. The helium cooling film is injected tangentially beneath a turbulent air boundary layer at an injection Mach number $\Mai=1.30$ and a total temperature ratio of $\TR=0.75$. The Mach number of the air freestream is $\Ma_\infty=2.45$. An undisturbed reference case and a case with shock interaction is investigated. The oblique shock is created by a flow deflection $\beta=8\deg$ and impinges upon the cooling film $53$ nozzle heights downstream of the injection location.

The experimental setup with the extension to helium cooling is described in section~\ref{sec:ExperimentalSetup}. The flow field is determined by high-speed PIV. Additional measurements are conducted by adding a highly dense seeding to the freestream, while the helium cooling film is injected without seeding. Consequently, the evaluation of the seeding density in the recorded images is used as a qualitative estimate of the helium fraction.
In section \ref{sec:Results}, the results are discussed. First, time averaged statistics of the flow field, i.e., the mean velocity field and the Reynolds shear stresses, of the undisturbed reference case and the shock interaction case are presented and compared to LES results of \cite{Konopka2013a}. Then, the turbulent transport of species, i.e., the distribution of the turbulent Schmidt number, is analyzed. Finally, the essential conclusions are drawn in section~\ref{sec:Conclusion}.
%

\section{Experimental setup}
\label{sec:ExperimentalSetup}

As stated above, the experimental setup used in \cite{Marquardt2019b} is extended to helium cooling flows. Therefore, the essential changes compared to the former analysis concern the helium fraction determination in subsection~\ref{sec:ParticleDensityBasedHeliumFractionEstimation} and the uncertainty discussion in subsection~\ref{sec:MeasurementUncertainties}.

\subsection{Wind tunnel and model}
\label{sec:WindTunnelAndModel}

All experiments were conducted in the trisonic wind tunnel which is an intermittently working vacuum storage tunnel in the Mach number range from 0.3 to 4.0. The flow is stable up to 3 seconds.
The unit Reynolds number varies between $6\cdot 10^6$ and $16\cdot10^6\,\mathrm{m}^{-1}$ depending on the Mach number and the ambient conditions.
The freestream Mach number $\Ma_\infty$ in the test section is calculated from the pressure ratio $p/p_0$.
The static pressure $p$ is measured via pressure taps in the test section side walls during the operation of the wind tunnel.
The total pressure $p_0$ is measured by the same transducer just before each test run.
The measurement error of the pressure transducer of $0.3\%$ full scale introduces an uncertainty in the Mach number determination of $\pm1.3\%$.

\begin{figure}[tb]
	\centering
		\includegraphics{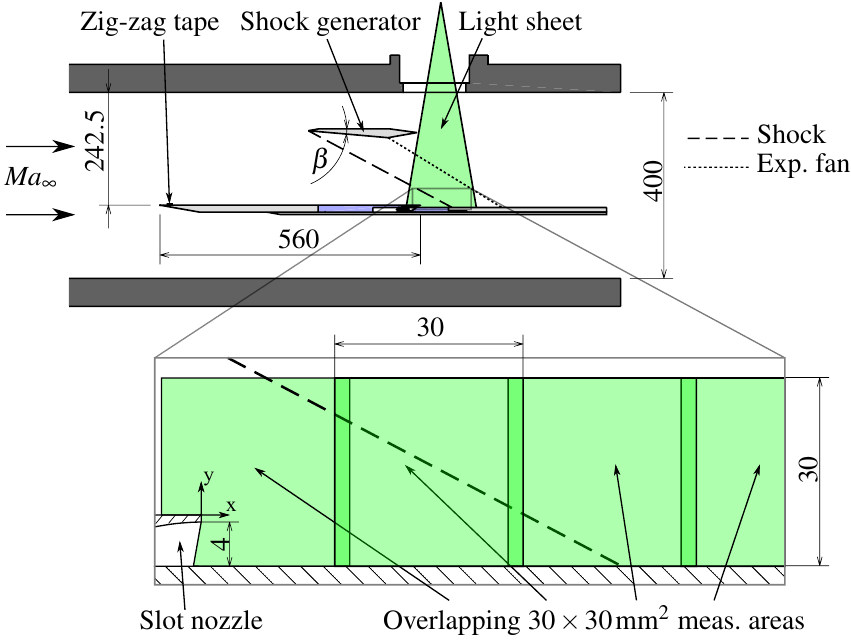}
	\caption{Dimensions of wind tunnel model and its location in the test section.}
	\label{fig:testSection}
\end{figure}

The model spans across the entire width of the $400\,\mm \times 400\,\mm$ test section of the wind tunnel. It has an overall length of $960\,\mm$ and possesses a thickness of $20\,\mm$. The dimensions of the model and its position in the test section are shown in figure~\ref{fig:testSection}. The flow is tripped by a $0.2\,\mm$ thick zig-zag tape $10\,\mm$ downstream of the wedge-shaped leading edge of the model to ensure a fully developed turbulent boundary layer at the position of injection. The cooling flow is injected $560\,\mm$ downstream of the leading edge through a $200\,\mm$ wide, centered slot nozzle and it develops on a $400\,\mm$ long flat plate. The cooling flow is channelized in the spanwise direction by two $2\,\mm$ thick and $20\,\mm$ high glass plates. To reduce the disturbances introduced by the glass plates, the leading edges exhibit an outward facing wedge shape. The oblique shock is generated by a wedge with an angle of $8\deg$ that spans the entire test section width.
The expansion fan emanating from the shock generator reaches the model $87\,\mm$ downstream of the shock impingement location.

\begin{figure}[tb]
	\centering
		\includegraphics{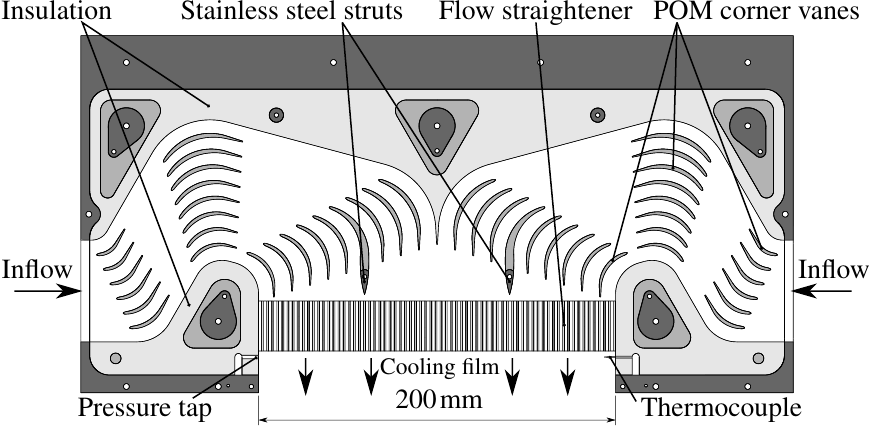}
	\caption{Spanwise cross section of the nozzle insert with plenum chamber.}
	\label{fig:PlenumCrossSection}
\end{figure}

An interchangeable nozzle insert enables to change the injection Mach number without dismounting the model from the test section. The nozzle insert is made of stainless steel with a thermal conductivity of $\SI{15}{\watt\per\meter\per\kelvin}$. It contains an insulated plenum chamber where the cooling flow is deflected to the main flow direction and homogenized by a flow straightener. A spanwise cross section of the nozzle insert with the plenum chamber is depicted in figure~\ref{fig:PlenumCrossSection}.

The helium flow enters the plenum chamber symmetrically through rectangular ducts at both sides of the model. Corner vanes inside the chamber guide the flow through the chamber and smoothly adapt the different cross-section areas between the inlets and the outlet to avoid flow separation inside the plenum. To reduce the heat flux between the cooling flow and the nozzle insert, the corner vanes are made of polyoxymethylene (POM) with a thermal conductivity of \SI{0.31}{\watt\per\meter\per\kelvin} and the material of the side walls of the air duct is rigid polyvinyl chloride (PVC) foam. Additionally, the top and bottom wall of the plenum chamber are insulated with a $1\,\mm$ thick layer of rigid PVC foam. The trailing edges of two corner vanes are part of the stainless steel nozzle insert and are fixed to the bottom wall of the model. This reduces the deformation of the nozzle due to the increased pressure inside the plenum chamber. Before the flow is accelerated to the injection Mach number, it runs through a honeycomb flow straightener with a cell size of $1.6\,\mm$ and a length of $28\,\mm$. Static pressure and temperature are monitored via a pressure tap and a thermocouple downstream of the flow straightener.

\begin{figure}[tb]
	\centering
		\includegraphics{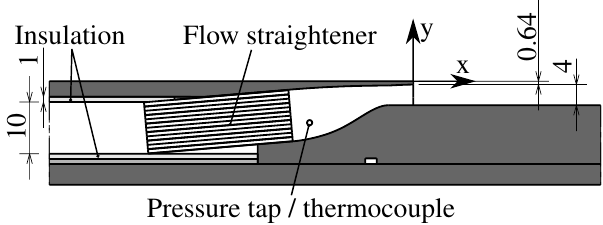}
	\caption{Cross section of the slot nozzle.}
	\label{fig:NozzleCrossSection}
\end{figure}

Figure~\ref{fig:NozzleCrossSection} shows a cross section of the nozzle. The flow straightener is inclined by an angle of approximately $5\deg$ to reduce the amount of tracer particles impacting on the bottom part of the nozzle. The supersonic part of the Laval nozzle is realized as the upper half of a symmetric, bell-shaped nozzle. The nozzle exit height is $S=4\,\mm$ and the thickness of the nozzle lip is $0.64\,\mm$.
To generate a steady cooling-film flow, the plenum chamber is fed with a constant mass flow such that the static pressure at the nozzle outlet equals the static pressure of the freestream in the wind tunnel. This constant mass flow is generated by a choked Venturi nozzle. The Venturi nozzle is fed from a bundle of twelve \SI{200}{\bar} helium cylinders. A pressure regulator upstream of the Venturi nozzle sets the pressure and, thus, the mass flow rate of the cooling flow. Downstream of the Venturi nozzle, the flow passes through two seeding generators with 6 Laskin nozzles each. A bypass controls the seeding density without changing the mass flow rate. The seeded cooling-film flow passes through a heat exchanger which precools the flow before it is split into two flows to enter the model symmetrically from both sides. The temperature is reduced to the final injection temperature in another pair of heat exchangers which are located close to the inlets of the nozzle insert.

The secondary sides of the heat exchangers are part of a closed-cycle cooling circuit filled with low viscosity polydimethylsiloxane (PDMS) oil. The cooling circuit is cooled by liquid nitrogen via an oil/liquid nitrogen heat exchanger. The liquid nitrogen flow rate is regulated by a closed-loop PID temperature controller to keep the temperature of the cooling circuit constant throughout the measurements. Depending on the injection Mach number and the ambient conditions, the cooling circuit is set to a temperature in the range of $\SI{-55}{\celsius}$ to $\SI{-40}{\celsius}$.

Between two measurements, a small helium mass flow is fed through the helium supply to keep the plenum chamber at low temperatures, which leads to a quick transient response to a steady temperature in the beginning of each measurement. With this setup, the mass flow rate and the injection temperature settle within $\SI{3}{s}$ to steady state. During the measurement time of $\SI{1.5}{s}$, the total temperature ratio $T_{0,\mathrm{i}}/T_{0,\infty}$ between the injection flow and the freestream flow is constant within $\pm0.2\%$. However, variations in the temperature of the cooling circuit lead to a fluctuation of the temperature between the measurements of up to $\pm1\%$.


\subsection{Particle-image velocimetry}
\label{sec:ParticleImageVelocimetry}

The particle-image velocimetry (PIV) setup consists of a Quantronix Darwin Duo 527-40-M laser and a Photron Fastcam SA5 high-speed PIV camera which are synchronized by an ILA synchronizer.
The light sheet enters the test section through a window in the ceiling. 
It is oriented vertically and parallel to the flow on the centerline of the model (figure~\ref{fig:testSection}). 
The thickness of the light sheet is $1\,\mm$.
The laser has an energy of $30\,\mathrm{mJ}$ per pulse. However, it is estimated that due to losses and the size of the field of view only approximately $11-15\,\mathrm{mJ}$ effectively illuminate the field of view.
The camera is mounted at a small angle $\approx2\deg$ to the normal of the light sheet under Scheimpflug condition to reduce aero-optical aberrations.
It is equipped with a $180\,\mm$ Tamron tele macro lens
at an aperture of $f/3.5$
 to realize a field of view of $30\,\mm \times 30\,\mm$ in the measurement plane.
The PIV system records 1000 samples per second with a resolution of $1024 \times 1024\,\mathrm{px}^2$.
To reduce the amount of laser light scattered from the model surface into the camera, the surface is highly polished.
The cooling flow as well as the main flow are seeded with Di-Ethyl-Hexyl-Sebacat (DEHS).
The seeding in the main flow is filtered using a cyclone particle separator that reduces the mean particle diameter. The particle response time is assessed in section~\ref{sec:MeasurementUncertainties}.

Each measurement consists of 1500 snapshots recorded over a measurement time of $1.5\,\mathrm{s}$.
First, to remove the background in the particle images, a sliding background is subtracted from the particle images. The sliding background is calculated for every laser cavity separately. An iterative procedure is used. First, the mean and the standard deviation of the image brightness is calculated over a span of 31 images for every pixel. In the following four iterations, values that deviate from the mean by more than $1.5$ times the standard deviation are excluded from the statistics. This effectively excludes pixels that show particles from the statistics and a much shorter span is sufficient to generate a convincing background image.
After the background subtraction, the particle images are preprocessed using a non-linear Gaussian blur to reduce camera noise and are dewarped using a camera calibration based on the Tsai model~(\cite{Tsai87}).
For the camera calibration, a CNC machined aluminum target with $\SI{0.3}{mm}$ holes filled with black paint with a spacing of $\SI{1}{mm}$ is used. Therefore, approx. $900$ calibration points are used for camera calibration.
To allow particle shifts larger than half the interrogation window size, the image evaluation uses a multi-grid approach with integer window shift to get an initial displacement field. Then, the velocity field is refined using an iterative predictor-corrector scheme with subpixel accurate image deformation according to the procedure described by \cite{Astarita2005}. The initial displacement is interpolated for each pixel of the image using a third-order B-Spline interpolation. Both images are deformed by half the displacement to get a second-order accurate estimate of the displacement field. The image interpolation uses Lanczos resampling, i.e., Lanczos windowed cardinal sine interpolation, incorporating the neighboring $8 \times 8\,\mathrm{px}^2$.
An integral velocity predictor is used to ensure convergence of the iterative scheme \cite{Scarano2004}. Hence, the predictor is the weighted average of the per-pixel displacement over the interrogation window. The corrector is determined by evaluating the cross-correlation function between both exposures with a 3-point Gaussian peak estimator \cite{Raffel2007}.
The initial window size for the multi-grid evaluation is $128 \times 128\,\mathrm{px}^2$ and the window size used for the iterative PIV evaluation is $32 \times 32\,\mathrm{px}^2$ with $75\%$ overlap corresponding to a physical size of $1 \times 1\,\mm^2$. This leads to a final vector pitch of $0.25\,\mm$ or $0.0625 S$. 
The windows of the iterative PIV evaluation are weighted by a Gaussian window with a standard deviation normalized by the window half width of $\sigma=0.5$.
Between the iterations, outliers in the vector field are detected using a normalized median test~\cite{Westerweel2005} and are replaced by interpolated values. A total of three multi-grid steps and five steps of the iterative evaluation are performed resulting in a validation rate over $90\%$ in the final dataset.
The surface reflections were masked in the recorded images. The first point used for PIV interrogation is at $\Delta y=0.25\,\mm$ off the wall.

Since the field of view is approximately $30 \times 30\,\mm^2$ in the current setup and the cooling-film flow evolves over a considerably greater length, the results for each set of flow parameters are composed of up to nine separate overlapping measurements along the centerplane of the model.
The bounds of each measurement are indicated by thin black lines in the final vector fields.

\subsection{Particle density based helium fraction estimation}
\label{sec:ParticleDensityBasedHeliumFractionEstimation}

To estimate the helium fraction in the flow, measurements with dense seeding in the freestream and no seeding in the helium cooling film are conducted. Thus, the distribution of the seeding density gives an estimate of the helium fraction. To generate the dense seeding in the freestream, a fog machine that vaporizes a glycol based fog fluid is used. This leads to a particle density of up to $15\,\mm^{-2}$ in the freestream. 
At some distance downstream of the injection, the helium cooling film and the air freestream have mixed sufficiently such that the seeding density in the cooling film allows the determination of the flow velocity via PIV. However, since the seeding density in the cooling film is still lower compared to the measurements with seeded cooling film, a physical windows size of $2 \times 2\,\mm^2$ is chosen for the evaluation of the measurements. The simultaneous evaluation of the flow velocity and seeding density, i.e., helium fraction, allows to estimate of the velocity-concentration correlation and the turbulent Schmidt number. It goes without saying that the results with full seeding and with seeding only in the cooling film were compared and found to agree well, especially concerning the separation bubble size and reattachment position.

To evaluate the particle density in the images, first, a sliding background is subtracted from the images as described in section~\ref{sec:ParticleImageVelocimetry}. 
The brightness threshold for the particle detection is based on the image noise level. Thus, the noise level of the images has to be estimated. For this, the algorithm of \cite{Immerkaer96} is used.
The images are normalized by the particle detection threshold which is set to the noise level in the present study.
Then, the laser light reflections on the ground are masked in the images. A Gaussian high-pass filter with a cutoff wavelength of $20\,\mathrm{px}$ is applied to remove the residual large scale gradients in the image. Additionally, to reduce the image noise, like in the post-processing for the PIV evaluation, a non-linear Gaussian blur is applied. For the non-linear Gaussian blur, the images are first processed by a gamma transform ($I_\mathrm{out}=I_\mathrm{in}^\gamma$) with $\gamma=0.1$ such that the dark pixels in the image that contain the noise are intensified. Then, a standard Gaussian blur with a cutoff wavelength of $3\,\mathrm{px}$ is applied. Finally, the image is converted back into the original linear scale by applying a gamma transform with $\gamma=10$. Note that negative values that might be present in the high-pass filtered images are set to zero during this procedure.
Then, the images are dewarped based on the same camera calibration as used in the PIV evaluation. Additionally, the images are deformed according to the velocity field. As in the PIV evaluation, the velocity field is projected onto each pixel by a third-order B-spline interpolation. Both exposures are shifted by half the particle shift using Lanczos resampling incorporating the neighboring $8 \times 8\,\mathrm{px}^2$. This effectively shifts the particle images to the same physical time for which the velocity is evaluated by the PIV algorithm.
Then, the deformed particle images of both exposures are multiplied. This helps to detect only valid particles and reject the image noise further. Particles which are located in the same location in both exposures constructively interfere, whereas particles that only occur in one exposure are reduced in brightness.
Finally, the particles are detected in the images.

To be considered as a particle, each pixel is tested for multiple conditions. First, since the images are normalized by the particle detection threshold, the brightness of the pixel must be larger than unity. Second, the brightness must be larger than the brightness of all $5 \times 5$ neighboring pixels. Third, a second-order two-dimensional polynomial is fitted to the $5 \times 5\,\mathrm{px}^2$ stencil around the current pixel. Only pixels for which the polynomial has negative curvature in both directions are considered a particle.
No further sub-pixel accurate estimate of the particle position is used.
To allow a PIV evaluation of the images without seeding in the cooling film, the physical interrogation window size for the concentration measurements is increased to $2 \times 2\,\mm^2$. This ensures a sufficient amount of tracer particles within the interrogation windows even in the cooling film. 
To match the spatial response of the particle density estimate to the velocity estimate, the particles are counted in regions with the same physical size and weighting function as in the PIV evaluation. 
Thus, the particles are counted within $2 \times 2\,\mm^2$ Gaussian weighted windows with a standard deviation normalized by the window half width of $\sigma=0.5$.

The air volume fraction $\fracAirVol$ is assumed to linearly scale with the particle density $\particleDensity$, i.e., $\fracAirVol=k \particleDensity$. Thus, the helium volume fraction is $\fracHeVol=1-\fracAirVol=1-k \particleDensity$ and the corresponding fluctuation is $\fracHeVol'=-k\particleDensity'$ with the unknown scaling factor $k$.
The scaling factor includes the unknown freestream particle density and the probability to successfully detect a particle which varies locally due to differences in the laser light intensity and image sharpness. In the results, where profiles of the turbulent mass transport are shown, the scaling factor $k$ of each distribution is determined from the freestream above the cooling film. 

The helium mass fraction $\fracHe$ is related to the helium volume fraction according to
\begin{equation}
\fracHe = \frac{\fracHeVol}{\left(1-\molarRatio\right)\fracHeVol+\molarRatio}
\label{eq:massVolFraction}
\end{equation}
with the ratio of the molar masses of helium and air $\molarRatio=\molarRatioDefinition\approx7.24$.
To obtain the helium mass fraction fluctuation, equation~\ref{eq:massVolFraction} is linearized around the mean helium volume fraction:
\begin{equation}
\fracHe'
=\left.\frac{\partial \fracHe}{\partial \fracHeVol}\right|_{\overline{\fracHeVol}}\fracHeVol'
=\frac{\molarRatio}{\left(\left(1-\molarRatio\right)\overline{\fracHeVol}+\molarRatio\right)^2}\fracHeVol'
\quad .
\label{eq:massFractionFluct}
\end{equation}
The turbulent Schmidt number 
\begin{equation}
\Sct=
\frac{\overline{\rho u' v'}}{\overline{\rho \fracHe' v'}}
\frac{\frac{\partial \overline{\fracHe}}{\partial y}}{\frac{\partial \overline{u}}{\partial y}}
\approx
\frac{\overline{\rho}\overline{u' v'}}{\overline{\rho}\overline{\fracHe' v'}}
\frac{\frac{\partial \overline{\fracHe}}{\partial y}}{\frac{\partial \overline{u}}{\partial y}}
=
\frac{\overline{u' v'}}{\overline{\fracHe' v'}}
\frac{\frac{\partial \overline{\fracHe}}{\partial y}}{\frac{\partial \overline{u}}{\partial y}}
\label{eq:TurbSchmidt}
\end{equation}
can be expressed in terms of the helium volume fraction
\begin{equation}
\Sct=
\frac{\overline{u' v'}}{\left.\frac{\partial\fracHe}{\partial \fracHeVol}\right|_{\overline{\fracHeVol}} \overline{\fracHeVol' v'}}
\frac{\overline{\frac{\partial \fracHe}{\partial \fracHeVol}\frac{\partial\fracHeVol}{\partial y}}}{\frac{\partial \overline{u}}{\partial y}}
\approx
\frac{\overline{u' v'}}{\overline{\fracHeVol' v'}}
\frac{\frac{\partial\overline{\fracHeVol}}{\partial y}}{\frac{\partial \overline{u}}{\partial y}}
\quad ,
\label{eq:TurbSchmidtVolume}
\end{equation}
since  
$\overline{\frac{\partial \fracHe}{\partial \fracHeVol}\frac{\partial\fracHeVol}{\partial y}}
\approx
\left.\frac{\partial \fracHe}{\partial \fracHeVol}\right|_{\overline{\fracHeVol}}\frac{\partial\overline{\fracHeVol}}{\partial y}$ is used for the Reynolds averaging.
Finally, the turbulent Schmidt number can be expressed in terms of the particle density and the scaling factor k which is assumed to be temporally invariant
\begin{equation}
\Sct=
\frac{\overline{u' v'}}{-k \overline{\particleDensity' v'}}
\frac{\frac{\partial \overline{\left(1-k\particleDensity \right)}}{\partial y}}{\frac{\partial \overline{u}}{\partial y}}
=
\frac{\overline{u' v'}}{-k\overline{\particleDensity' v'}}
\frac{-k \frac{\partial \overline{\particleDensity}}{\partial y} - \particleDensity \frac{\partial k}{\partial y}}{\frac{\partial \overline{u}}{\partial y}}
\quad .
\label{eq:TurbSchmidt2}
\end{equation}
Thus, for a sufficiently small wall-normal gradient of the scaling factor ${\partial k}/{\partial y}$, the turbulent Schmidt number is independent of the scaling factor
\begin{equation}
\Sct=
\frac{\overline{u' v'}}{\overline{\particleDensity' v'}}
\frac{\frac{\partial \overline{\particleDensity}}{\partial y}}{\frac{\partial \overline{u}}{\partial y}}
\quad .
\label{eq:TurbSchmidt3}
\end{equation}
In the present measurements, the scaling factor changes locally due to the inhomogeneous laser intensity within the light sheet. However, the light sheet is oriented approximately normal to the wall. Therefore, the laser intensity mostly changes in the streamwise direction, while being approximately constant in the wall-normal direction. Thus, neglecting the wall-normal gradient of the scaling factor is justified in the present measurements.

\subsection{Measurement uncertainties}
\label{sec:MeasurementUncertainties}

To estimate the flow tracking capability of the DEHS and the smoke particles, measurements for a flow encountering an oblique shock with a deflection angle of $\beta=5\deg$ were performed at a freestream Mach number $\Ma_\infty=2.45$. The analysis of the flow field across the abrupt velocity change allows the determination of the particle relaxation length, and, hence, the effective particle diameter. This effective particle diameter is used to calculate the particle response times for arbitrary flow conditions.
The procedure to determine the particle relaxation time and length is described in detail in \cite{Marquardt2019a}.

Three kinds of seeding generators are used. For the pure velocity measurements, the DEHS seeding generator for the freestream uses a cyclone separator to reduce the particle size, whereas a seeding generator without filter was used for the cooling film. 
The effective particle diameter for the seeding generator without filter is $d_\mathrm{p}=1.19\,\um\pm4\%$. For the seeding generator with filter, it is $d_\mathrm{p}=0.7\,\um\pm9\%$.
For the helium fraction estimation, a smoke generator was used to achieve a sufficiently high seeding density. The effective particle diameter of the smoke generator is $d_\mathrm{p}=1.24\,\um\pm8\%$.
Depending on the flow conditions and the flow medium, the particles show different relaxation times and lengths which are calculated for small velocity changes assuming Stokes flow as summarized in table~\ref{tab:SummaryOfParticleCharacteristics}. The uncertainty of the relaxation time $\relaxTime$ and length $\relaxLength$ is in the range $\pm18\%$ for the filtered DEHS particles, $\pm7\%$ for the unfiltered DEHS particles, and $\pm16\%$ for the smoke particles. Throughout the paper, a confidence level of $95\%$ is used for all uncertainties.
\begin{table}[htb]
	\caption{Summary of particle characteristics.}
	\label{tab:SummaryOfParticleCharacteristics}
	\centering
\begin{tabular}{cccccc}
\multicolumn{1}{l}{} & \multicolumn{1}{l}{} & \multicolumn{2}{c}{DEHS particles} & \multicolumn{2}{c}{smoke particles} \\ \cline{3-6} 
 & \multicolumn{1}{l}{} & \begin{tabular}[c]{@{}c@{}}freestream\\ (air)\\ filtered particles\end{tabular} & \begin{tabular}[c]{@{}c@{}}cooling film\\ (helium)\\ unfiltered particles\end{tabular} & \begin{tabular}[c]{@{}c@{}}freestream\\ (air)\end{tabular} & \begin{tabular}[c]{@{}c@{}}cooling film\\ (helium)\end{tabular} \\ \hline
Mach number & $[-]$ & $2.45$ & $1.30$ & $2.45$ & $1.30$ \\
$\relaxTime$ & $[\us]$ & $2.9$ & $4.8$ & $6.4$ & $6.9$ \\
$\relaxLength$ & $[\mm]$ & $1.7$ & $1.7$ & $5.8$ & $6.2$ \\ \hline
\end{tabular}
\end{table}

\cite{Samimy91} investigated the motion of tracer particles in a compressible free shear layer. The velocity error grows approximately linearly with the Stokes number $\St=\relaxTime/{\tau_\mathrm{f}}$, i.e., the ratio of the particle relaxation time $\relaxTime$ and the flow time scale ${\tau_\mathrm{f}}$, meaning approximately $2\%$ error for $\St=0.2$. For flow visualizations, however, the authors recommended a Stokes number $\St\leq0.05$. The flow time scale $\tau_\mathrm{f}$ in their study is defined as $\tau_\mathrm{f}=10 \delta_{\omega 0}/\left(u_1-u_2\right)$ with $\delta_{\omega 0}$ being the vorticity thickness. In the field of view of the present measurement, i.e., at $\xs\geq36$, the maximum velocity difference between the shear layer and the cooling film is $u_1-u_2 \approx 164\,\si{m/s}$ and the minimum vorticity thickness is $\delta_{\omega 0}\approx 4.3\,\mm$ which results in a maximum Stokes number of $\St=0.026$. Closer to the injection nozzle, however, the Stokes number increases due to the larger velocity difference and smaller shear layer thickness. At $\xs=20$, the Stokes number is $\St=0.057$, at $\xs=10$ it is $\St=0.1$ and at $\xs=5$ it reaches a value of $\St=0.18$. Thus, the tracer particles do not track the flow accurately close to the injection. Therefore, some particles can be transported from the air freestream into the helium cooling film. This introduces systematical errors in the measurements further downstream since the helium species is assumed to have zero particle density.

The small field of view and the high velocities of up to $900\,\mathrm{m/s}$ require pulse distances of $1000\,\ns$.
Hence, the relatively long laser pulse width of $210\,\ns$ introduces a significant amount of particle blur in the recorded particle images.
Additionally, due to slight differences in the temporal pulse shape of both laser cavities, the effective pulse distance differs by $\pm 40\,\ns$ from the set pulse distance.
This systematic error, which can be as high as $4\%$ in the current measurements, has to be accounted for.
To reduce this error, each measurement is conducted twice, where the cavities are triggered in reverse order between the measurements.
Then, the systematic error occurs with opposite sign in both measurements such that it can be determined and a corrected pulse distance can be used to calculate the velocity fields. A typical value of the uncertainty in the corrected pulse distance is in the range of $2-4\,\mathrm{ns}$. This results in an uncertainty of less than $1\%$ in the velocity. This approach is described in detail in \cite{Marquardt2019a}.

Another source of systematic uncertainty is introduced by the determination of the freestream velocity $u_\infty$ which is calculated from the static to total pressure ratio and the total temperature of the wind tunnel. Since $u_\infty$ is used in the normalization, the measurement uncertainty of the pressure sensors and the temperature probe accumulate to an uncertainty of $1.2\%$ for quantities normalized by $u_\infty$ and $2.4\%$ for quantities normalized by $u_\infty^2$.

Statistical uncertainties arise from the limited amount of samples. This uncertainty is calculated for the mean value as well as for higher-order statistics according to the variance estimates given by \citet{Benedict96}. For cases with strong separation, the uncertainty of the mean velocity reaches values of up to $1.2\%$ of the freestream velocity $u_\infty$.
All uncertainties, including the uncertainties in the determination of the Mach number and the total temperature, are combined using the square root of the sum of the squared quantities assuming uncorrelated error sources. 

The counting of randomly distributed particles within an evaluation window is assumed to resemble a Poisson distribution. Therefore, the variance of the particle count $\particleCount$ within an evaluation window is expected to be $\mathit{var}\left(\particleCount\right)=\particleCount$ even at a statistically constant particle density. At particle counts on the order of $\particleCount=20$ this leads to a standard deviation of the detected particle count of approximately $22\%$. While the mean particle count can be evaluated to the desired accuracy by increasing the number of snapshots, the rms particle count is contaminated with a measurement noise of at least $\sqrt{\particleCount}$. The Poisson noise is assumed to be uncorrelated to the flow velocity. Therefore, even though the velocity-concentration correlation is reduced to some extent by the addition of uncorrelated measurement noise, it is supposed to resemble the qualitative trends in the flow field.

\section{Results}
\label{sec:Results}

The freestream Mach number is $\Ma_\infty=2.45$ and the nozzle height $S$ based freestream Reynolds number is $\Res=40,500$. 
The total temperature and total pressure of the freestream are ambient conditions which are in the range $290\,\mathrm{K} \leq T_{0,\infty} \leq 300\,\mathrm{K}$ and $979\,\mathrm{hPa} \leq p_{0,\infty} \leq 1002\,\mathrm{hPa}$.
 Cooled helium is injected underneath a turbulent boundary layer at an injection Mach number $\Mai=1.30$. The total temperature ratio between the freestream and the cooling-film flow is $\TR=0.75$. The static pressure of the injected flow matches the freestream condition. The Reynolds number of the cooling flow is $\Rei=7,009$.
A flow deflection of $\beta=8\deg$ generates an oblique shock wave that impinges upon the cooling film. The theoretical shock impingement location, i.e., the extrapolation of the shock orientation and location above the shear layer onto the wall, is $\xsimp=53.0$. 

Note that the LES results of \cite{Konopka2013a} that are used for comparison were conducted at $\ResLES=13,500$ and $\Ma_{\infty,\mathrm{LES}}=2.44$. The helium cooling film was injected at a Mach number $\MaiLES=1.30$ and an injection Reynolds number $\ReiLES=2,561$. The shock wave that was generated by a flow deflection of $\beta=8\deg$ impinges upon the wall at $\mathit{x}_{\mathrm{imp},\mathrm{LES}}/S=52.1$.

All flow parameters of this study are summarized in table~\ref{tab:FlowParameters}.
The experimental cases are denoted as case~\caseExpRef{} for the no-shock reference case and case~\caseExpShock{} for the flow with shock interaction. Likewise, the numerical cases are referred to as cases~\caseNumRef{} and \caseNumShock{}.
Since the unit Reynolds number of the wind tunnel can not be controlled, the Reynolds number of the experiments is determined solely by the size of the wind tunnel model. A model size that is large enough to allow high resolution measurements of the cooling film is chosen. Hence, the Reynolds number of the experiments is higher by a factor of three compared to the LES Reynolds number. However, the model was designed such that the non-dimensional boundary layer thickness ${\delta_{99}/S}$ at the injection location of the experiment and the simulation is comparable. Furthermore, the experiments and the simulations agree in terms of the freestream Mach number $\Ma_\infty$, the model geometry, i.e., the thickness of the nozzle lip, the injection parameters, i.e., Mach number $\Mai$, total temperature ratio $\TR$, and blowing rate $M$, as well as the shock deflection angle $\beta$ and the pressure ratio across the shock $p_2/p_1$. The experimental shock impingement position $\xsimp$ only slightly deviates from the numerical location. These differences between the numerical and the experimental flow problem, which are primarily determined by the Reynolds number, mean that no perfect quantitative but a qualitative agreement of the LES and the measurement data can be expected.
 
%

In the following, first, the flow field is analyzed in terms of the mean velocity field and the Reynolds shear stress in subsection~\ref{sec:MeanFlow}. The measurements of the no-shock case and the shock interaction case are compared to the LES results to show the experimental and numerical agreement. Then, the turbulent mass transport is analyzed in subsection~\ref{sec:TurbulentTransport}. The streamwise and the wall-normal turbulent mass flux as well as the turbulent Schmidt number are discussed and compared to the simulation findings.

\begin{table}[htb]
\caption{Flow parameters, i.e., freestream Mach number ${\Ma_\infty}$, freestream Reynolds number ${\Res}$, non-dimensional boundary layer thickness ${\delta_{99}/S}$, injection Mach number ${\Mai}$, injection Reynolds number ${\Rei}$, injection total temperature ratio ${\TR}$, blowing rate ${M=\frac{\rho_\mathrm{i} u_\mathrm{i}}{\rho_\infty u_\infty}}$, non-dimensional shock impingement position ${\xsimp}$, deflection angle ${\beta}$, shock angle ${\sigma}$, and static pressure ratio across the shock ${p_2/p_1}$.}
\label{tab:FlowParameters}
\centering
\begin{tabular}{lccccccccccc}
\hline\\[-2ex]
\multicolumn{1}{c}{Case} & $\Ma_\infty$ & \multicolumn{1}{l}{$\Res$} & \multicolumn{1}{l}{$\frac{\delta_{99}}{S}$} & \multicolumn{1}{l}{$\Mai$} & \multicolumn{1}{l}{$\Rei$} & \multicolumn{1}{l}{$\frac{T_{0,\mathrm{i}}}{T_{0,\infty}}$} & \multicolumn{1}{l}{$M$} & $\frac{\ximp}{S}$ & $\beta [\deg]$ & $\sigma [\deg]$ & $\frac{p_2}{p_1}$ \\[1ex] \hline
\caseExpRef{} & $2.45$ & $40,500$ & $2.15$ & $1.30$ & $7,009$ & $0.75$ & $0.2$ & - & - & - & - \\
\caseExpShock{} & $2.45$ & $40,500$ & $2.15$ & $1.30$ & $7,009$ & $0.75$ & $0.2$ & $53.0$ & $8$ & $30.6$ & $1.64$ \\
\caseNumRef{} & $2.44$ & $13,500$ & $2.27$ & $1.30$ & $2,561$ & $0.75$ & $0.2$ & - & - & - & - \\
\caseNumShock{} & $2.44$ & $13,500$ & $2.27$ & $1.30$ & $2,561$ & $0.75$ & $0.2$ & $52.1$ & $8$ & $30.7$ & $1.64$
\end{tabular}
\end{table}

\subsection{Mean flow field}
\label{sec:MeanFlow}

\subsubsection*{Flow without shock interaction}

\begin{figure}[b]
	\centering
		\includegraphics{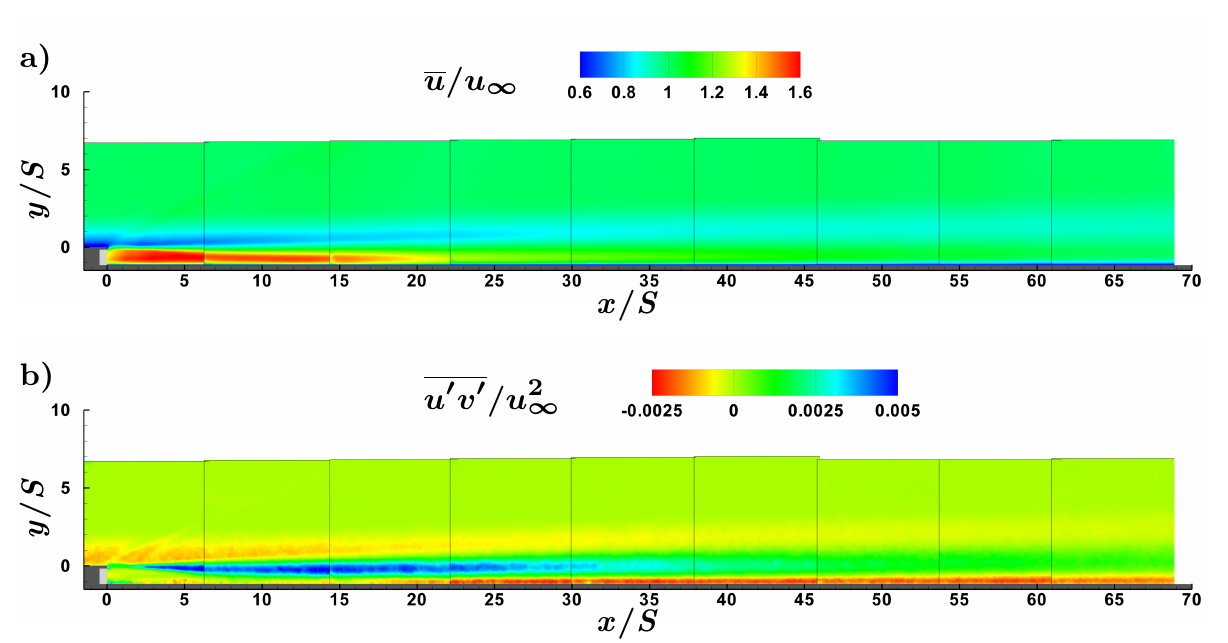}
	\caption{Mean streamwise velocity $\u$ and Reynolds shear stress $\uv$ for case~\caseExpRef{}.}
	\label{fig:PIV_noShock}
\end{figure}

In figure~\ref{fig:PIV_noShock}, the mean streamwise velocity $\u$ and the Reynolds shear stress $\uv$ are shown for the undisturbed, i.e., no-shock, reference case~\caseExpRef{}. 
Due to the nominal injection velocity of $\u=1.6$, the velocity in the cooling film exceeds the freestream velocity up to approximately $\xs=30$ downstream of the injection nozzle. This results in a negative wall-normal velocity gradient in the mixing layer which leads to a positive Reynolds shear stress in the mixing layer. The velocity deficit of the shear layer is apparent in the entire field of view of the measurements. Therefore, even though the flow relaxes towards the end of the field of view, it does not resemble a boundary-layer-like flow but shows the characteristics of the wall-jet region.
In the vicinity of the bottom wall, negative values of the Reynolds shear stress indicate the development of a turbulent boundary layer.

\begin{figure}[h]
	\centering
		\includegraphics{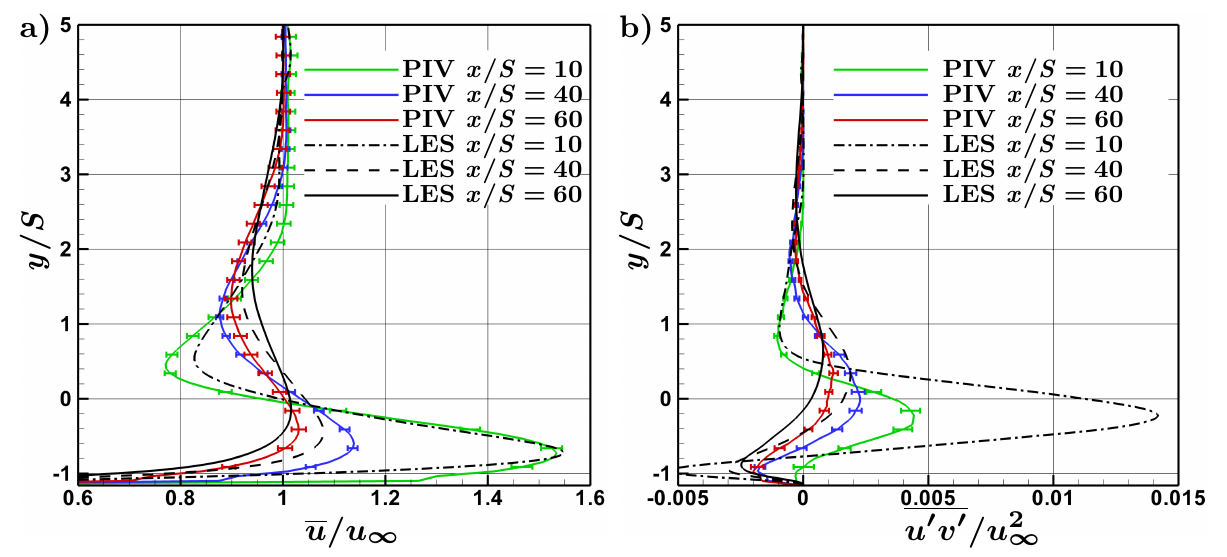}
	\caption{Profiles of the mean streamwise velocity $\u$ and the Reynolds shear stress $\uv$ for case~\caseExpRef{} and for case~\caseNumRef{}.}
	\label{fig:PIV_noShock_profiles}
\end{figure}

A comparison of the mean streamwise velocity $\u$ and the Reynolds shear stress $\uv$ of the present PIV measurements and the LES results of \cite{Konopka2013a} is depicted in figure~\ref{fig:PIV_noShock_profiles} for the no-shock case~\caseExpRef{}. The wall-normal profiles are extracted at three streamwise positions, i.e., $\xs=10$, $\xs=40$, and $\xs=60$.
At $\xs=10$, the mean velocity distributions of the measurements and the simulation are in good agreement. That is, the extrema, their locations, and the general shape of the profiles correspond. However, the Reynolds shear stress differs significantly. The LES shows a peak Reynolds shear stress in the mixing layer that exceeds the measurement by a factor of approximately three. Additionally, there is a discrepancy in the near-wall region. 
The deviation in the near-wall region can be attributed to the spatial response of the PIV measurements. 
Due to the small boundary layer thickness on the order of the interrogation window size, the near-wall velocity distribution is definitely not fully resolved by the PIV measurements.

Regarding the profiles at $\xs=40$ and $\xs=60$, the stronger turbulent mixing in the LES leads to a more intense relaxation of the flow further downstream. The peak velocity in the cooling film and the maximum Reynolds shear stress in the mixing layer decrease somewhat faster and the minimum velocity in the shear layer increases a bit quicker in the LES results compared to the present measurements. At $\xs=40$, the peak velocity in the cooling film exceeds the numerical results by $5.5\%$ and at $\xs=60$ the velocity is higher by $1.5\%$. The Reynolds shear stress in the mixing layer decreases slower in the experiments. This leads to higher values of the Reynolds shear stress in the experiments. At $\xs=40$, the peak Reynolds shear stress in the mixing layer is higher by $\Delta \uv=0.0004$ and at $\xs=60$ it is higher by $\Delta \uv=0.0002$. Especially with respect to the Reynolds shear stress distributions, the differences in the experimental and numerical Reynolds number have to be kept in mind. That is, higher shear stress values are to be expected in the experiments. The higher Reynolds number of the experiments leads to less pronounced thickening of the boundary layer such that the peak velocity in the cooling film and the peak Reynolds shear stress in the mixing layer are located closer to the wall. Nevertheless, it can be stated that for such a multiple shear layer interaction the qualitative agreement of the experimental and numerical distributions is satisfactory.

\subsubsection*{Flow with shock interaction}

\begin{figure}[p]
	\centering
		\includegraphics{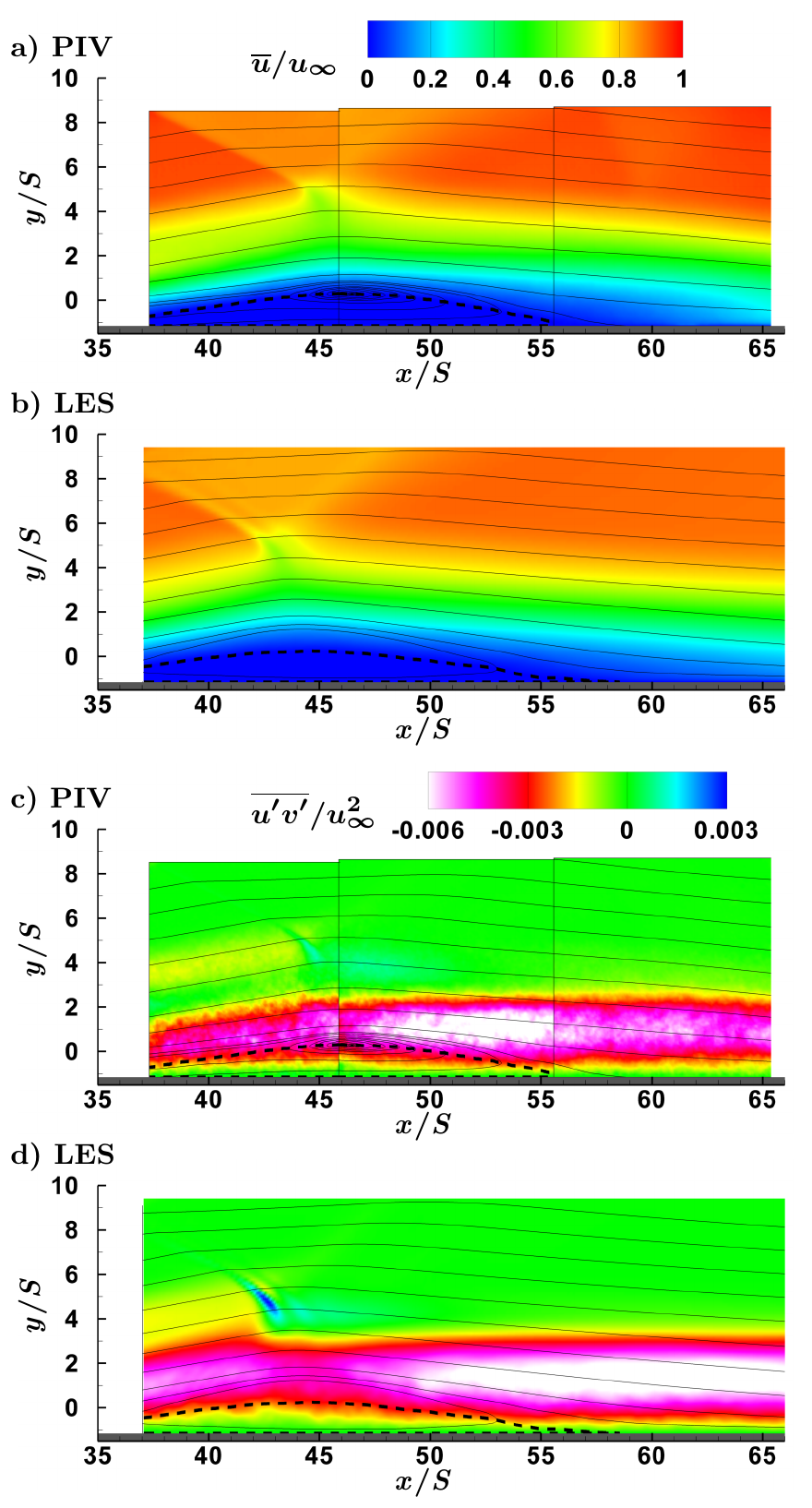}
	\caption{Mean streamwise velocity $\u$ and Reynolds shear stress $\uv$ distribution for case~\caseExpShock{} and for case~\caseNumShock{}.}
	\label{fig:PIV_LES_shock}
\end{figure}

In figure~\ref{fig:PIV_LES_shock}, the distribution of the mean streamwise velocity $\u$ and the Reynolds shear stress $\uv$ are shown for case~\caseExpShock{}, i.e., the flow configuration with shock interaction, and for the corresponding case~\caseNumShock{} of \cite{Konopka2013a}. 
In the experimental and the numerical results, the impinging shock generates a large separation bubble. Note that due to limited optical access to the test section, the experimental results do not cover the separation point. The dashed line in figure~\ref{fig:PIV_LES_shock} indicates zero mean streamwise velocity. When the separation bubble thickness is expressed in terms of the dividing streamline, the thickness differs between the LES and the measurements. The thickness based on the dividing streamline is $\Delta \ys=2.0$ in the PIV results and $\Delta \ys=2.5$ in the LES. Reattachment occurs at $\xs=56$ in the PIV results, whereas in the LES, a very thin layer of mean back flow close to the wall extends further downstream to $\xs=59$. If this very thin backflow region is not considered, the reattachment is located at $\xs=56$. Due to the slight difference in the theoretical shock impingement position, i.e., $\xsimp=53$ in the experiments and $\xsimp=52.1$ in the simulation, the position of maximum separation bubble thickness and the position where the impinging shock penetrates the shear layer are located about $2\,S$ further downstream in the PIV results. Considering the shift of the impinging shock and the thinner wall-normal extent of the backflow region at the start of the field of view of the measurements, it is expected that the separation point in the experiments is located further downstream compared to the numerical result. Hence, the separation bubble is slightly smaller in the experiments which is also consistent with the higher experimental Reynolds number.

The shock induced separation bubble leads to intense turbulent mixing. The distributions of the Reynolds shear stress in figures~\ref{fig:PIV_LES_shock}~c) and \ref{fig:PIV_LES_shock}~d) indicate a strong turbulent transport towards the wall. The peak of the Reynolds shear stress is located downstream of the separation bubble approximately $\Delta \ys=2.4$ off the wall. Although the peak value of the Reynolds shear stress is similar in the measurements and the simulations, the peak is located close to the separation bubble in the experiments and decays quickly further downstream. In the simulation, the peak is located approximately $10\,S$ further downstream and the decay is much less pronounced.

\begin{figure}[h!]
	\centering
		\includegraphics{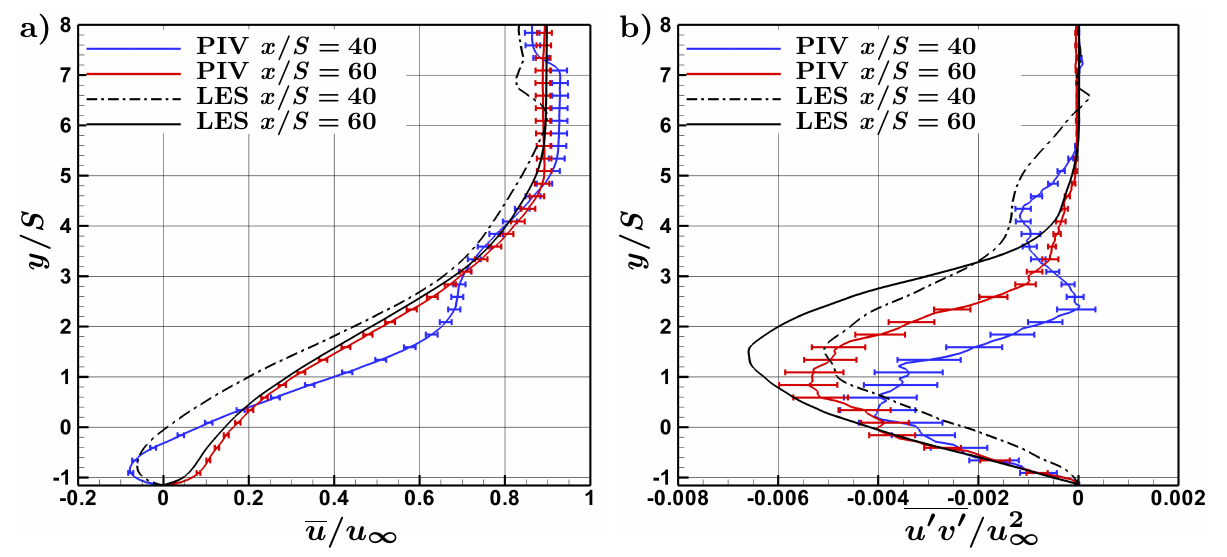}
	\caption{Profiles of the mean streamwise velocity $\u$ and the Reynolds shear stress $\uv$ for the shock interaction case~\caseExpShock{} and for case~\caseNumShock{}.}
	\label{fig:PIV_LES_shock_profiles}
\end{figure}

In figure~\ref{fig:PIV_LES_shock_profiles}, profiles of the mean streamwise velocity $\u$ and the Reynolds shear stress $\uv$ are shown for the shock interaction case~\caseExpShock{} and for the LES case~\caseNumShock{}. The profiles are determined at two streamwise positions, i.e., within the separation bubble at $\xs=40$ and downstream of the reattachment position at $\xs=60$.
The impinging shock passes through the profile at $\xs=40$ at $\ys=6.4$ for the LES and at $\ys=7.0$ for the PIV measurement. This deviation is a result of the slight difference in the shock impingement location.
Due to the slightly larger separation bubble in the LES, the shear layer is located further off the wall in the numerical results at $\xs=40$. The upper edge of the shear layer can be identified in the Reynolds shear stress profile at $\ys=6.3$ in the LES results and at $\ys=5.5$ in the measurements.
There is a velocity plateau in the range $2\leq \ys \leq 3$ in the measurements but not in the simulation. This plateau stems from the cooling film that is deflected around the separation bubble. In the LES, however, the stronger turbulent mixing and faster relaxation of the flow that has been discussed earlier makes the cooling film indistinguishable in the profiles.
Downstream of the separation bubble, i.e., at $\xs=60$, the mean velocity profiles of the LES and the PIV measurement are in good agreement. Due to the intense turbulent mixing, the cooling film has merged with the shear layer and is neither apparent in the mean velocity profile, nor in the Reynolds shear stress profile of the measurement.
The smaller separation bubble size of the experiment caused by the higher Reynolds number leads to a turbulent mixing zone that is located approximately $1\,S$ closer to the wall with a peak value that is approximately $18\%$ smaller than in the simulation.

In conclusion, keeping in mind the higher experimental Reynolds number, the experimental results show that the mixing between the cooling film and the freestream is lower compared to the numerical findings. Hence, the flow relaxes slower and the features of the film cooling flow, i.e., a cooling film and a free shear layer separated by a mixing layer, persist further downstream. A thinner boundary layer develops on the bottom wall which shifts the locations of the peak velocity and the peak Reynolds shear stress closer to the wall.
With shock interaction, the separation bubble in the experiment is slightly smaller than in the LES. This can be attributed to the higher near-wall momentum of the cooling film. Due to the smaller relaxation of the flow, the velocity in the cooling film upstream of the separation bubble is approximately $5\%$ higher. In the experiments, the cooling film and the shear layer are still distinguishable when the flow is deflected by the separation bubble, whereas in the numerical data both layers have already merged at the shock interaction location due to the faster relaxation of the flow. Downstream of the separation bubble, strong turbulent mixing occurs. Due to the smaller separation bubble size, the region of intense mixing downstream of the separation bubble is located closer to the wall in the experiments and the peak value is smaller.

\subsection{Turbulent mass transport}
\label{sec:TurbulentTransport}

\subsubsection*{Flow without shock interaction}

To estimate the turbulent mass transport, measurements have been conducted without any seeding injected into the cooling film. Thus, the seeding in the freestream acts as a tracer for the species air which in turn allows to estimate the helium mass fraction $\fracHe$ as discussed in section~\ref{sec:ParticleDensityBasedHeliumFractionEstimation}.

\begin{figure}[b]
	\centering
		\includegraphics{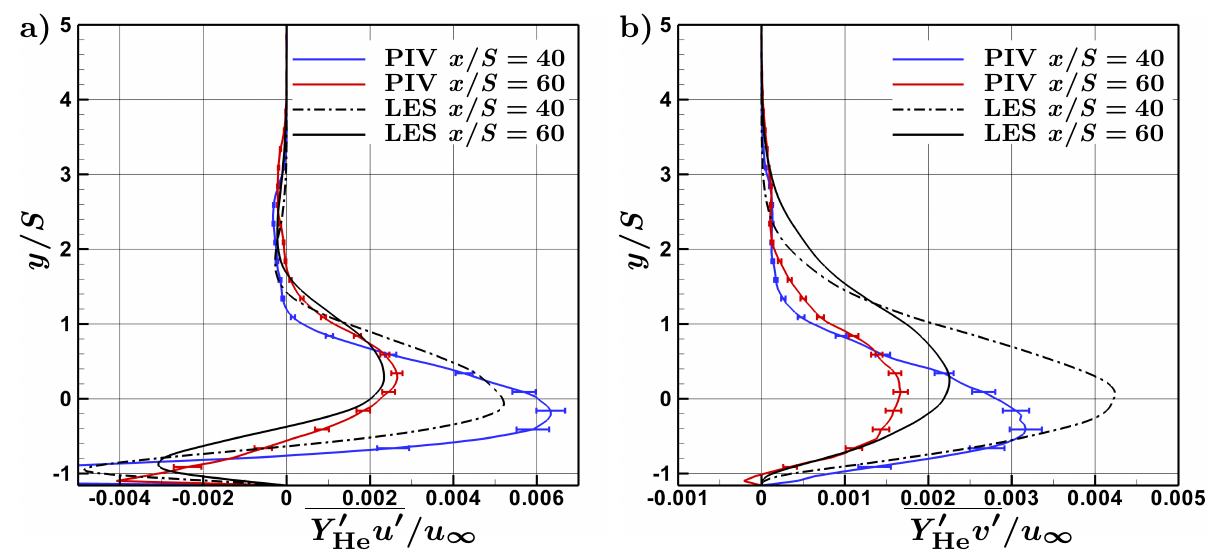}
	\caption{Streamwise and wall-normal turbulent flux $\fracHeU$ and $\fracHeV$ for case~\caseExpRef{} and for case~\caseNumRef{}.}
	\label{fig:PIV_LES_noShock_turbulentFlux_profiles}
\end{figure}

In figure~\ref{fig:PIV_LES_noShock_turbulentFlux_profiles}, the profiles of the streamwise and wall-normal turbulent flux $\fracHeU$ and $\fracHeV$ are shown for case~\caseExpRef{} in comparison to the corresponding LES results of \cite{Konopka2013a} at two streamwise positions, i.e., at $\xs=40$ and $\xs=60$. The error bars in the plots indicate the statistical error in the determination of the turbulent fluxes. The systematical error due to the uncertainty of the scaling factor $k$ that would lead to a different scaling of the turbulent fluxes is not included.
The profiles of the streamwise turbulent flux $\fracHeU$ in figure~\ref{fig:PIV_LES_noShock_turbulentFlux_profiles}~a) resemble the numerical data. A slightly negative correlation is found in the shear layer, i.e., in the range $1 \leq \ys \leq 3.5$, at both positions. The mixing layer is characterized by a positive turbulent flux $\fracHeU$. Close to the wall, the turbulent flux shows a negative peak. The near-wall peak in the profile at $\xs=40$ is chopped off. As discussed earlier, the spatial averaging of the evaluation influences the results and additional errors are introduced by laser reflections close to the wall that might be mistaken for particles in the estimate of the particle density.
The peak flux in the mixing layer decreases along the streamwise direction. At $\xs=60$, the maximum value decreases to approximately half the value at $\xs=40$ in the simulation and the measurements.

The wall-normal turbulent flux $\fracHeV$ is shown in figure~\ref{fig:PIV_LES_noShock_turbulentFlux_profiles}~b). The results show a stretched positive extremum from the wall into the shear layer. Thus, the flux is directed off the wall. The maxima are located in the mixing layer at the same height as the corresponding positive peaks of the streamwise flux in figure~\ref{fig:PIV_LES_noShock_turbulentFlux_profiles}~a). Due to the thinner boundary layer in the experiments, the maxima are located slightly closer to the wall compared to the LES. The maximum values, however, disagree between the LES and the present data. The estimated peak value of the measurements is approximately $25\%$ lower. The aforementioned scaling factor $k$ affects the streamwise and the wall-normal flux equally. Thus, it is apparent that the measurements and the simulation quantitatively deviate in the ratio between the streamwise and the wall-normal turbulent mass flux. 
An explanation might be found in the noise characteristics of the PIV measurements. The wall-normal velocity fluctuations $v'$ are smaller than the streamwise velocity fluctuations $u'$. However, both are subject to the same amount of measurement noise. Thus, the wall-normal correlation $\fracHeV$ suffers from a larger relative amount of uncorrelated noise that tends to reduce the correlation between the velocity and the concentration. Nevertheless, the LES and the measurements consistently show a decrease of the peak value of approximately $50\%$ between $\xs=40$ and $\xs=60$.

\begin{figure}[b]
	\centering
		\includegraphics{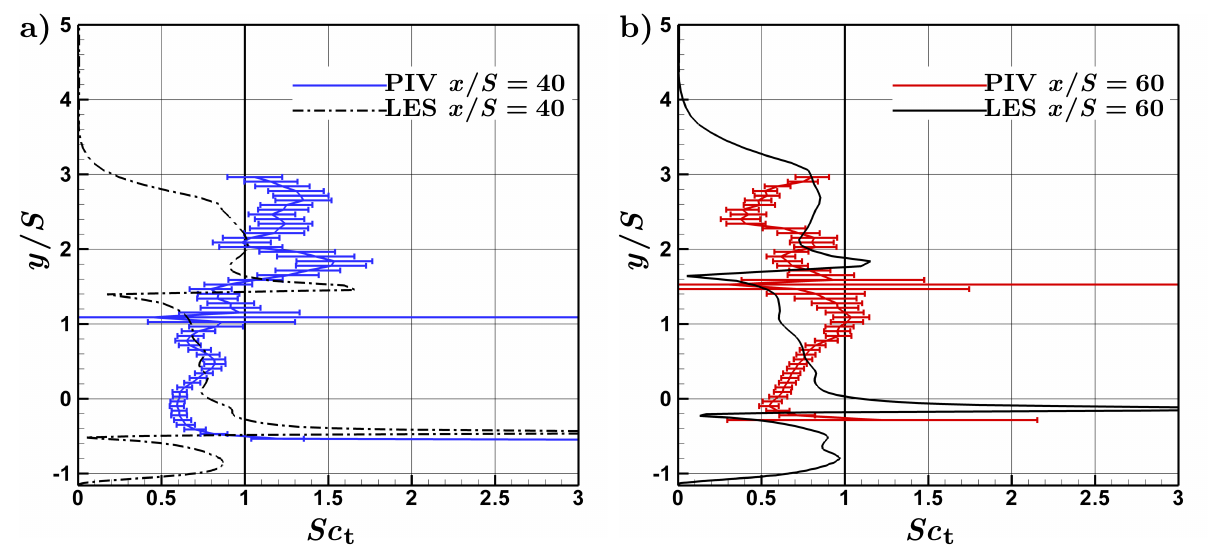}
	\caption{Profiles of the turbulent Schmidt number $\Sct$ at two streamwise positions for case~\caseExpRef{} and for case~\caseNumRef{}.}
	\label{fig:PIV_LES_noShock_Sct_profiles}
\end{figure}

As described in detail in section~\ref{sec:ParticleDensityBasedHeliumFractionEstimation}, the turbulent Schmidt number can be derived from the measurements. The profiles of the turbulent Schmidt number are shown in figure~\ref{fig:PIV_LES_noShock_Sct_profiles} for case~\caseExpRef{} and for the LES case~\caseNumRef{} at two streamwise positions, i.e., $\xs=40$ and $\xs=60$. Regions in which the measurement uncertainty dominates the results, i.e., in the freestream with a diminishing wall-normal gradient of the streamwise velocity and close to the wall where $\overline{\fracHe'v'}$ gets small, are excluded from the illustration. The error bars indicate the statistical uncertainties of all terms in equation~\ref{eq:TurbSchmidt3}. However, bias errors that are introduced by neglecting the gradient of the unknown scaling factor $k$ or by the linearization are not included.
Note that a vanishing mean velocity gradient $\partial \overline{u}/\partial y$ causes discontinuities in the profiles of the turbulent Schmidt number. Since the boundary layer is thinner in the experiments, the zero crossings, and hence, the discontinuities are located closer to the wall in the measurements.
The determination of the turbulent Schmidt number based on the particle density shows, although the shape of the experimental and numerical profiles partially disagree, a reasonable agreement with the LES results for the range of the turbulent Schmidt number that occurs in the flow field. Within the mixing layer and the shear layer, the turbulent Schmidt number is predominantly in the range $0.5 \leq \Sct \leq 1.5$. Like in the numerical analysis the mean turbulent Schmidt number is lower at $\xs=60$ than at $\xs=40$.

The results of the turbulent fluxes and the turbulent Schmidt number show that their determination based on the particle density distribution allows not only a qualitative interpretation but also gives reasonable quantitative results. The mass eddy diffusivity is determined accurate enough to reasonably quantify the turbulent Schmidt number, i.e., the ratio between the eddy viscosity and the mass eddy diffusivity. 
Nevertheless, there are differences in the profiles of the turbulent Schmidt number of the simulation and the experiments.
The deviations could be caused by neglecting the wall-normal gradient of the scaling factor $k$. This factor includes the probability to successfully detect particles in the evaluation. It mainly depends on the local signal to noise ratio of the particles. Variations of the laser intensity within the field of view influences the particle brightness and thus the signal to noise ratio. On the other hand, the noise level might vary locally depending on the camera sensor. The image background level also causes local variations in the noise level. Optical blur due to the camera lens causes particles to appear dimmer and broader which also introduces local variations in the particle image-to-noise ratio.

\subsubsection*{Flow with shock interaction}

\begin{figure}[h!]
	\centering
		\includegraphics{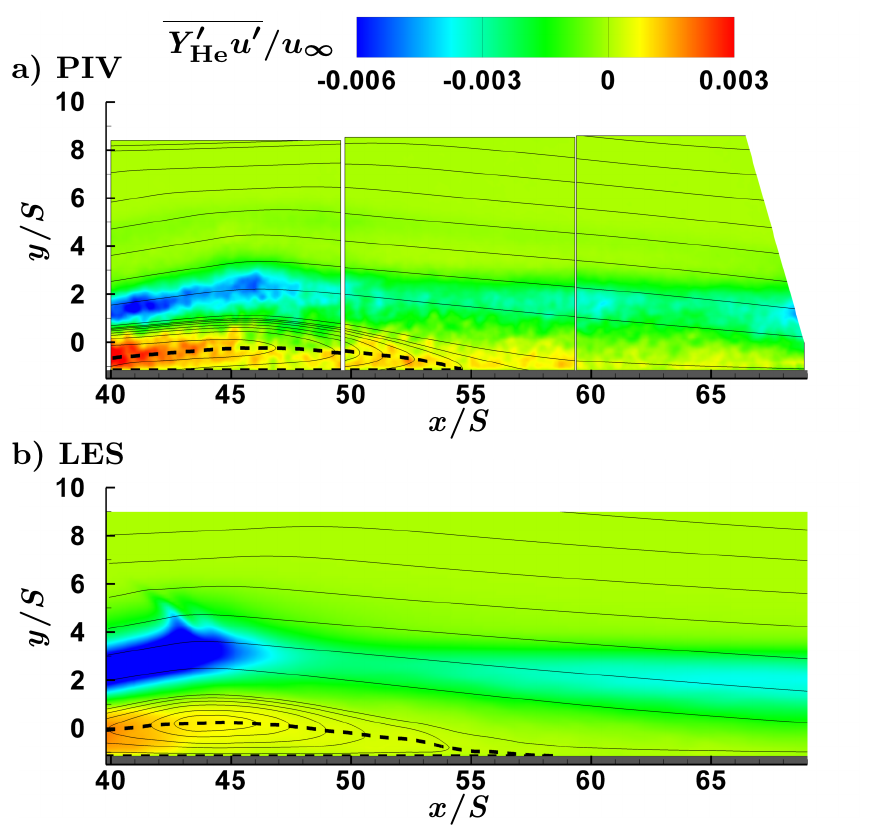}
	\caption{Streamwise turbulent mass flux $\fracHeU$ for case~\caseExpShock{} and for case~\caseNumShock{}.}
	\label{fig:PIV_LES_shock_HeU}
\end{figure}

The discussion of the flow field in section~\ref{sec:MeanFlow} showed the flow field with shock interaction to be considerably more complex than without shock interaction. This is also confirmed by the qualitative distributions of the turbulent fluxes and the turbulent Schmidt number which are considered next.

Figure~\ref{fig:PIV_LES_shock_HeU} shows the distribution of the streamwise turbulent mass flux $\fracHeU$ for the shock interaction case~\caseExpShock{} and for the corresponding LES case~\caseNumShock{}.
The distributions of the measurements and the simulation qualitatively agree. Due to the higher Reynolds number, the shock impinges slightly further downstream and the separation bubble is slightly thinner in the experiments. Within the separation bubble, the streamwise turbulent flux $\fracHeU$ is positive. 
Outside the separation bubble, the streamwise flux is highly negative upstream of the maximum bubble thickness. Further downstream, the streamwise flux remains negative but the magnitude is greatly reduced.

\begin{figure}[b]
	\centering
		\includegraphics{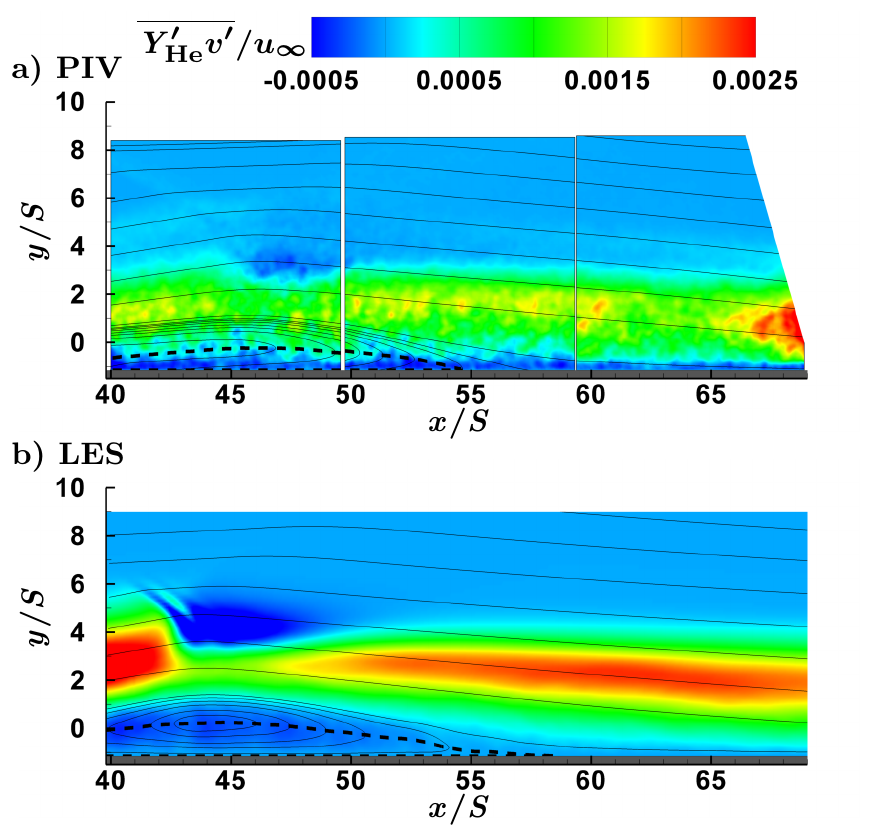}
	\caption{Wall-normal turbulent mass flux  $\fracHeV$ for case~\caseExpShock{} and for case~\caseNumShock{}.}
	\label{fig:PIV_LES_shock_HeV}
\end{figure}

The wall-normal flux $\fracHeV$ is shown in figure~\ref{fig:PIV_LES_shock_HeV}. It is positive within the shear layer outside the separation bubble upstream of the shock impingement position. In the shock impingement region, the numerical and the experimental results show a region with negative wall-normal turbulent flux in the shear layer. It is located approximately in the range $\xs=46-50$ at a height of  $\ys=3$ in the experiments and in the range $\xs=44-50$ and $\ys=4$ in the simulation. Further downstream, the shear layer exhibits a distinct region of positive flux that is elongated in the streamwise direction.
In general, the magnitude of the wall-normal flux is smaller in the experiments. Additionally, the wall-normal flux in the separation bubble differs. While the LES shows a small negative flux within the whole separation bubble, the more slender separation bubble in the experiments possesses a positive flux in a region very close to the wall.

\begin{figure}[h]
	\centering
		\includegraphics{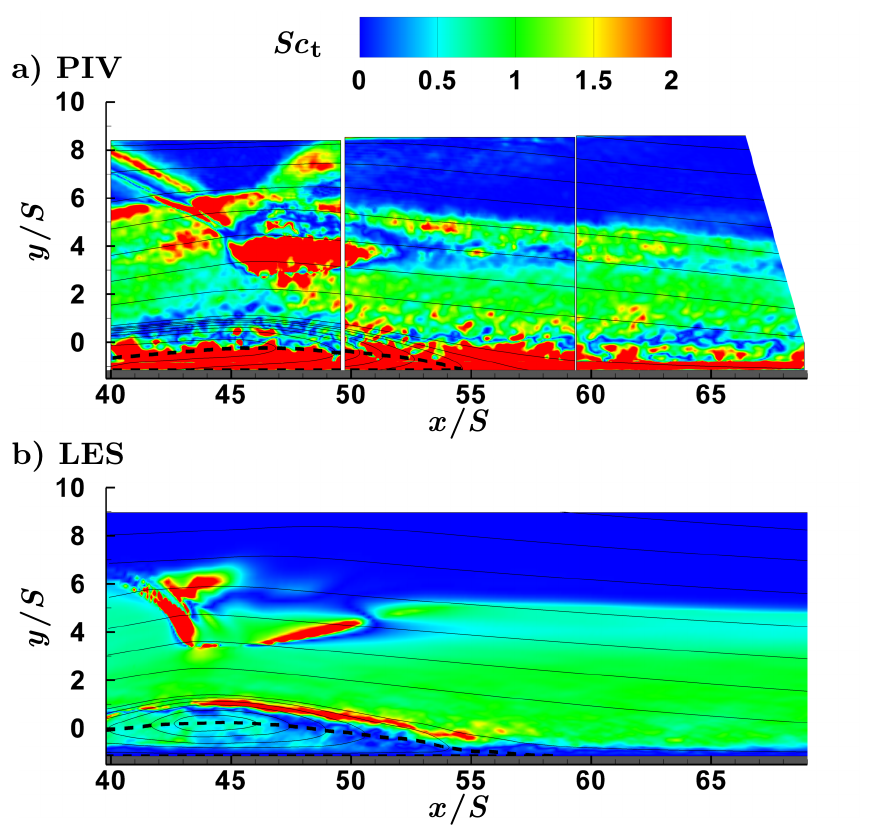}
	\caption{Turbulent Schmidt number $\Sct$ for case~\caseExpShock{} and for case~\caseNumShock{}.}
	\label{fig:PIV_LES_shock_Sct}
\end{figure}

The contours of the turbulent Schmidt number for case~\caseExpShock{} and for the LES case~\caseNumShock{} are illustrated in figure~\ref{fig:PIV_LES_shock_Sct}.
The turbulent Schmidt number distributions of the LES exhibit several distinct features that are qualitatively captured by the measurements.
When the shear layer passes through the incident shock, the turbulent Schmidt number increases significantly above unity in the experimental and numerical data. Closer to the wall, a region is located that starts at the foot of the impinging shock and is constrained by low values of the turbulent Schmidt number. While in the LES the turbulent Schmidt number in this region is only very high at the foot of the shock and at the end of this region, the turbulent Schmidt number is increased in the entire region in the measurements. Further downstream, i.e., approximately at $\xs=51$ and $\ys=4$, another region of low turbulent Schmidt number is located followed by a growing turbulent Schmidt number in the streamwise direction. This development in the shear layer can be identified in the LES results and in the measurements.
In the near-wall region, however, both results deviate which is related to the growing uncertainty of the experimentally determined turbulent Schmidt number in the sublayer.

\begin{figure}[h]
	\centering
		\includegraphics{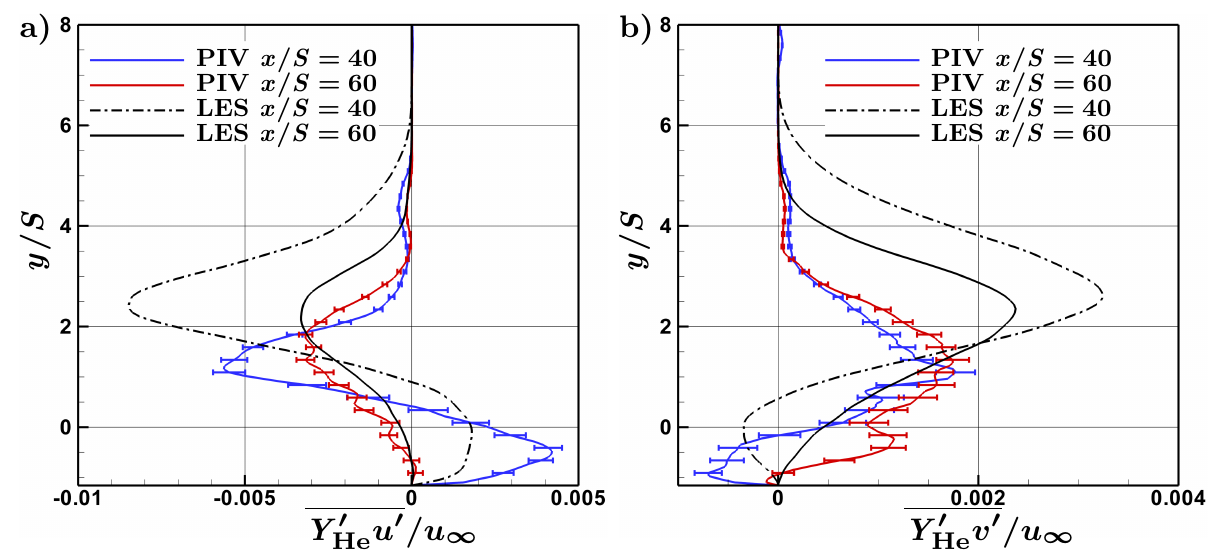}
	\caption{Profiles of the streamwise and wall-normal turbulent mass flux $\fracHeU$ and $\fracHeV$ at two streamwise positions for case~\caseExpShock{} and for case~\caseNumShock{}.}
	\label{fig:PIV_LES_shock_turbulentFlux_profiles}
\end{figure}

Profiles of the streamwise and the wall-normal turbulent mass flux $\fracHeU$ and $\fracHeV$ are depicted in figure~\ref{fig:PIV_LES_shock_turbulentFlux_profiles} for the shock interaction case~\caseExpShock{} and for the LES case~\caseNumShock{} at two streamwise positions. The first position is located within the separation bubble upstream of the maximum thickness of the separation bubble, i.e., at $\xs=40$, and the second profile is determined at $\xs=60$, i.e., downstream of the reattachment location.
At the upstream position $\xs=40$, the turbulent fluxes of the measurement match the qualitative shape of the numerical results. Due to the higher Reynolds number and the thinner separation bubble in the experiments, the peaks are located closer to the wall. 
The streamwise turbulent flux $\fracHeU$ is negative in the separated cooling film and positive in the separation bubble. The wall-normal flux $\fracHeV$ is positive in the separated cooling film and negative in the separation bubble. The peaks of the streamwise and the wall-normal turbulent flux are located approximately at the same normal distance above the wall.
Due to the Reynolds number difference, the quantitative values of the numerical and experimental data do not agree. For instance, the peaks in the separation bubble are approximately twice as high in the experiments.
Downstream of the reattachment position, i.e., in the profile at $\xs=60$, the streamwise turbulent flux exhibits a single negative peak in the numerical and the experimental results. In the experiments, the peak is located approximately $1\,S$ closer to the wall. The wall-normal turbulent flux is positive at $\xs=60$. Thus, the turbulent transport of helium is directed off the wall.

\begin{figure}[h]
	\centering
		\includegraphics{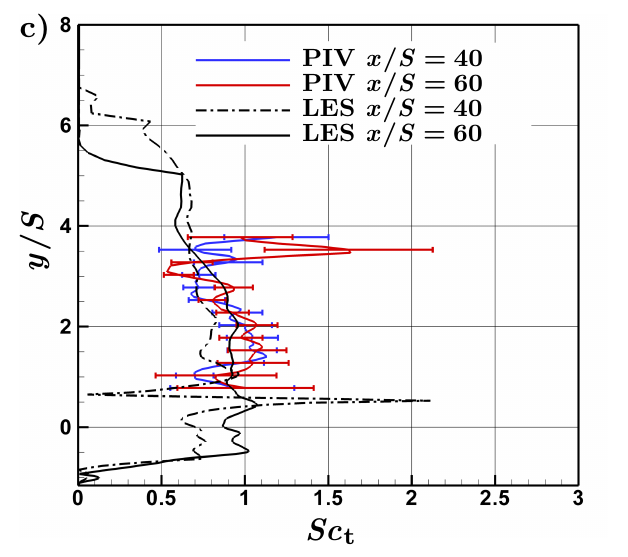}
	\caption{Profiles of the turbulent Schmidt number $\Sct$ at two streamwise positions for case~\caseExpShock{} and for case~\caseNumShock{}.}
	\label{fig:PIV_LES_shock_Sct_profiles}
\end{figure}

Profiles of the turbulent Schmidt number at both streamwise locations are shown in figure~\ref{fig:PIV_LES_shock_Sct_profiles}. Only the region where the determination of the turbulent Schmidt number is not dominated by measurement uncertainty is shown. Thus, the distribution of the turbulent Schmidt number is illustrated in the range $1 \leq\ys\leq 4$. The turbulent Schmidt number is in agreement with the numerical results in $1 \leq\ys\leq 3$. In this region, the turbulent Schmidt number of the LES drops from approximately $\Sct=0.9$ close to the wall to $\Sct=0.7$. The experimental data follow the LES within the measurement uncertainty.

\section{Conclusion}
\label{sec:Conclusion}

The interaction of a shock impinging on a helium cooling film was investigated using 2C high-speed PIV and measurements of the particle distribution.
The helium cooling film was injected tangentially at a Mach number $\Mai=1.30$ underneath a turbulent air boundary layer at a freestream Mach number $\Ma_\infty=2.45$. The temperature of the injected air was lowered to obtain a total temperature ratio between the freestream and the cooling film of $\TR=0.75$. An oblique shock was generated by a flow deflection of $\beta=8\deg$. A shock interaction case and a reference case without shock interaction are investigated. Large-eddy simulation (LES) results from \citet{Konopka2013a} are used for comparison. Except for the Reynolds number, which is three times higher in the experimental study, the geometry and the flow parameters match the simulations.

In addition to the high-speed PIV measurements, where seeding was added to the freestream and the cooling film, measurements without seeding in the cooling film were conducted. Based on the evaluation of the particle density in the recorded images the air volume fraction and the helium mass fraction in the flow were determined. The field of view of the measurements is located downstream of the injection such that the cooling film has partially mixed with the air freestream. Therefore, the particle density in the cooling film is sufficiently high to allow a simultaneous PIV evaluation of the recorded images. The velocity-concentration correlations, i.e., the streamwise and wall-normal turbulent mass flux $\fracHeU$ and $\fracHeV$, could be calculated and the turbulent Schmidt number $\Sct$ was determined.

The high speed of sound of the helium leads to a nominal injection velocity of $u_i/u_\infty=1.6$. This results in a negative velocity gradient in the mixing layer which causes positive values of the Reynolds shear stress, i.e., turbulent transport of momentum off the wall. Consequently, the velocity of the cooling film quickly decays.
When a shock impinges on the cooling film, a large separation bubble is generated and the turbulent mixing downstream of the bubble is strongly increased.
Due to the higher Reynolds number in the experiments, the wall boundary layer is thinner which shifts the mixing layer and the shear layer closer to the wall. In addition, the mixing layer grows slower, which leads to a slower decay of the velocity of the cooling film in the experiments. At shock interaction, the higher near-wall velocity in the experiments causes the separation bubble to be smaller compared to the simulation. Thus, the values of the Reynolds shear stress downstream of the separation bubble are smaller in the experiments compared to the simulation.
Nevertheless, the experimental distributions physically match the output of the LES findings despite the different Reynolds numbers.

The major focus of this study was on the measurements and analysis of the turbulent transport of mass in terms of the streamwise and the wall-normal turbulent mass fluxes and the turbulent Schmidt number. In general, the comparison with the LES from \citet{Konopka2013a} shows a good qualitative agreement of the turbulent mass fluxes. Discrepancies in the wall-normal distance of the maxima of the turbulent mass fluxes are due to the differences of the flow field caused by the varying Reynolds numbers. The distribution of the turbulent mass flux $\fracHeV$ shows that in the mixing layer helium is transported off the wall and air is transported into the cooling film. With shock interaction, however, in a small region downstream of the foot of the shock the direction of the wall-normal turbulent mass flux is reversed. 

In agreement with the simulations, the experiments clearly show variations of the turbulent Schmidt number within the flow field. With shock interaction, the variations are quite drastic, i.e., the turbulent Schmidt number changes strongly within a thin layer. The present experimental study confirms that for accurately predicting the turbulent mass flux, the assumption of a constant Schmidt number is inadequate. Not only for a film cooling configuration with shock interaction, the variation of the turbulent Schmidt number must be considered to obtain reasonable predictions of the turbulent mass flux, and hence, the cooling fluid concentration and the cooling effectiveness, which are essential for an efficient film cooling design.

\begin{acknowledgements}
This research was funded by the Deutsche Forschungsgemeinschaft within the research project ``Experimental Investigation of Turbulent Supersonic Film-Cooling Flows'' (SCHR 309/62-1).
\end{acknowledgements}

%
%

\bibliography{references}   

%
%

\end{document}